\documentclass[10pt, journal, final]{IEEEtran}

\usepackage[draft]{hyperref}  
\usepackage{gensymb}
\usepackage{amsmath,amssymb,amsfonts}
\usepackage{epsfig}
\usepackage{cite}
\usepackage{hhline}
\usepackage{multirow}
\usepackage{framed}
\usepackage{xcolor}
\usepackage{lipsum}
\usepackage{enumerate}
\usepackage{color}
\usepackage{graphicx}
\usepackage{subfigure}
\usepackage{hyperref}
\usepackage{algorithmic}
\usepackage[ruled]{algorithm}
\usepackage{booktabs}
\usepackage{bm}
\usepackage{tikz}
\usetikzlibrary{automata,arrows,positioning,calc}

\begin{document}

\title{The Transitional Behavior of Interference in \\ Millimeter Wave Networks and Its Impact on Medium Access Control}
\author{Hossein Shokri-Ghadikolaei,~\IEEEmembership{Student Member,~IEEE} and Carlo Fischione,~\IEEEmembership{Member,~IEEE}%
\thanks{The authors are with KTH Royal Institute of Technology, Stockholm,
Sweden (email: \{hshokri, carlofi\}@kth.se).}
\thanks{This work was supported by the Swedish Research Council under the project ``In-Network Optimization''.}
}


\newtheorem{defin}{Definition}
\newtheorem{theorem}{Theorem}
\newtheorem{prop}{Proposition}
\newtheorem{lemma}{Lemma}
\newtheorem{corollary}{Corollary}
\newtheorem{alg}{Algorithm}
\newtheorem{remark}{Remark}
\newtheorem{result}{Result}
\newtheorem{example}{Example}
\newtheorem{notations}{Notations}
\newtheorem{assumption}{Assumption}

\newcommand{\be}{\begin{equation}}
\newcommand{\ee}{\end{equation}}
\newcommand{\ba}{\begin{array}}
\newcommand{\ea}{\end{array}}
\newcommand{\bea}{\begin{eqnarray}}
\newcommand{\eea}{\end{eqnarray}}
\newcommand{\combin}[2]{\ensuremath{ \left( \ba{c} #1 \\ #2 \ea \right) }}
\newcommand{\diag}{{\mbox{diag}}}
\newcommand{\rank}{{\mbox{rank}}}
\newcommand{\dom}{{\mbox{dom{\color{white!100!black}.}}}}
\newcommand{\range}{{\mbox{range{\color{white!100!black}.}}}}
\newcommand{\image}{{\mbox{image{\color{white!100!black}.}}}}
\newcommand{\herm}{^{\mbox{\scriptsize H}}}  
\newcommand{\sherm}{^{\mbox{\tiny H}}}       
\newcommand{\tran}{^{\mbox{\scriptsize T}}}  
\newcommand{\tranIn}{^{\mbox{-\scriptsize T}}}  
\newcommand{\card}{{\mbox{\textbf{card}}}}
\newcommand{\asign}{{\mbox{$\colon\hspace{-2mm}=\hspace{1mm}$}}}
\newcommand{\ssum}[1]{\mathop{ \textstyle{\sum}}_{#1}}

\newcommand{\vbar}{\raisebox{.17ex}{\rule{.04em}{1.35ex}}}
\newcommand{\vbarind}{\raisebox{.01ex}{\rule{.04em}{1.1ex}}}
\newcommand{\D}{\ifmmode {\rm I}\hspace{-.2em}{\rm D} \else ${\rm I}\hspace{-.2em}{\rm D}$ \fi}
\newcommand{\T}{\ifmmode {\rm I}\hspace{-.2em}{\rm T} \else ${\rm I}\hspace{-.2em}{\rm T}$ \fi}
\newcommand{\B}{\ifmmode {\rm I}\hspace{-.2em}{\rm B} \else \mbox{${\rm I}\hspace{-.2em}{\rm B}$} \fi}
\newcommand{\Hil}{\ifmmode {\rm I}\hspace{-.2em}{\rm H} \else \mbox{${\rm I}\hspace{-.2em}{\rm H}$} \fi}
\newcommand{\C}{\ifmmode \hspace{.2em}\vbar\hspace{-.31em}{\rm C} \else \mbox{$\hspace{.2em}\vbar\hspace{-.31em}{\rm C}$} \fi}
\newcommand{\Cind}{\ifmmode \hspace{.2em}\vbarind\hspace{-.25em}{\rm C} \else \mbox{$\hspace{.2em}\vbarind\hspace{-.25em}{\rm C}$} \fi}
\newcommand{\Q}{\ifmmode \hspace{.2em}\vbar\hspace{-.31em}{\rm Q} \else \mbox{$\hspace{.2em}\vbar\hspace{-.31em}{\rm Q}$} \fi}
\newcommand{\Z}{\ifmmode {\rm Z}\hspace{-.28em}{\rm Z} \else ${\rm Z}\hspace{-.38em}{\rm Z}$ \fi}

\newcommand{\sgn}{\mbox {sgn}}
\newcommand{\var}{\mbox {var}}
\newcommand{\E}{\mbox {E}}
\newcommand{\cov}{\mbox {cov}}
\renewcommand{\Re}{\mbox {Re}}
\renewcommand{\Im}{\mbox {Im}}
\newcommand{\cum}{\mbox {cum}}

\renewcommand{\vec}[1]{{\bf{#1}}}     
\newcommand{\vecsc}[1]{\mbox {\boldmath \scriptsize $#1$}}     
\newcommand{\itvec}[1]{\mbox {\boldmath $#1$}}
\newcommand{\itvecsc}[1]{\mbox {\boldmath $\scriptstyle #1$}}
\newcommand{\gvec}[1]{\mbox{\boldmath $#1$}}

\newcommand{\balpha}{\mbox {\boldmath $\alpha$}}
\newcommand{\bbeta}{\mbox {\boldmath $\beta$}}
\newcommand{\bgamma}{\mbox {\boldmath $\gamma$}}
\newcommand{\bdelta}{\mbox {\boldmath $\delta$}}
\newcommand{\bepsilon}{\mbox {\boldmath $\epsilon$}}
\newcommand{\bvarepsilon}{\mbox {\boldmath $\varepsilon$}}
\newcommand{\bzeta}{\mbox {\boldmath $\zeta$}}
\newcommand{\boldeta}{\mbox {\boldmath $\eta$}}
\newcommand{\btheta}{\mbox {\boldmath $\theta$}}
\newcommand{\bvartheta}{\mbox {\boldmath $\vartheta$}}
\newcommand{\biota}{\mbox {\boldmath $\iota$}}
\newcommand{\blambda}{\mbox {\boldmath $\lambda$}}
\newcommand{\bmu}{\mbox {\boldmath $\mu$}}
\newcommand{\bnu}{\mbox {\boldmath $\nu$}}
\newcommand{\bxi}{\mbox {\boldmath $\xi$}}
\newcommand{\bpi}{\mbox {\boldmath $\pi$}}
\newcommand{\bvarpi}{\mbox {\boldmath $\varpi$}}
\newcommand{\brho}{\mbox {\boldmath $\rho$}}
\newcommand{\bvarrho}{\mbox {\boldmath $\varrho$}}
\newcommand{\bsigma}{\mbox {\boldmath $\sigma$}}
\newcommand{\bvarsigma}{\mbox {\boldmath $\varsigma$}}
\newcommand{\btau}{\mbox {\boldmath $\tau$}}
\newcommand{\bupsilon}{\mbox {\boldmath $\upsilon$}}
\newcommand{\bphi}{\mbox {\boldmath $\phi$}}
\newcommand{\bvarphi}{\mbox {\boldmath $\varphi$}}
\newcommand{\bchi}{\mbox {\boldmath $\chi$}}
\newcommand{\bpsi}{\mbox {\boldmath $\psi$}}
\newcommand{\bomega}{\mbox {\boldmath $\omega$}}

\newcommand{\bolda}{\mbox {\boldmath $a$}}
\newcommand{\bb}{\mbox {\boldmath $b$}}
\newcommand{\bc}{\mbox {\boldmath $c$}}
\newcommand{\bd}{\mbox {\boldmath $d$}}
\newcommand{\bolde}{\mbox {\boldmath $e$}}
\newcommand{\boldf}{\mbox {\boldmath $f$}}
\newcommand{\bg}{\mbox {\boldmath $g$}}
\newcommand{\bh}{\mbox {\boldmath $h$}}
\newcommand{\bp}{\mbox {\boldmath $p$}}
\newcommand{\bq}{\mbox {\boldmath $q$}}
\newcommand{\br}{\mbox {\boldmath $r$}}
\newcommand{\bs}{\mbox {\boldmath $s$}}
\newcommand{\bt}{\mbox {\boldmath $t$}}
\newcommand{\bu}{\mbox {\boldmath $u$}}
\newcommand{\bv}{\mbox {\boldmath $v$}}
\newcommand{\bw}{\mbox {\boldmath $w$}}
\newcommand{\bx}{\mbox {\boldmath $x$}}
\newcommand{\by}{\mbox {\boldmath $y$}}
\newcommand{\bz}{\mbox {\boldmath $z$}}

\newenvironment{Ex}
{\begin{adjustwidth}{0.04\linewidth}{0cm}
\begingroup\small
\vspace{-1.0em}
\raisebox{-.2em}{\rule{\linewidth}{0.3pt}}
\begin{example}
}
{
\end{example}
\vspace{-5mm}
\rule{\linewidth}{0.3pt}
\endgroup
\end{adjustwidth}}

\newcommand{\Hossein}[1]{{\textcolor{blue}{\emph{**Hossein: #1**}}}}
\newcommand{\Challenge}[1]{{\textcolor{red}{#1}}}
\newcommand{\NEW}[1]{{\textcolor{blue}{#1}}}


\maketitle

\begin{abstract}
Millimeter wave (mmWave) communication systems use large number of antenna elements that can potentially overcome severe channel attenuation by narrow beamforming. Narrow-beam operation in mmWave networks also reduces multiuser interference, introducing the concept of noise-limited wireless networks as opposed to interference-limited ones. The noise-limited or interference-limited regime heavily reflects on the medium access control (MAC) layer throughput and on proper resource allocation and interference management strategies. Yet, these regimes are ignored in current approaches to mmWave MAC layer design, with the potential disastrous consequences on the communication performance. In this paper, we investigate these regimes in terms of collision probability and throughput. We derive tractable closed-form expressions for the collision probability and MAC layer throughput of mmWave ad~hoc networks, operating under slotted ALOHA. The new analysis reveals that mmWave networks may exhibit a non-negligible transitional behavior from a noise-limited regime to an interference-limited one, depending on the density of the transmitters, density and size of obstacles, transmission probability, operating beamwidth, and transmission power. Such transitional behavior necessitates a new framework of adaptive hybrid resource allocation procedure, containing both contention-based and contention-free phases with on-demand realization of the contention-free phase.
Moreover, the conventional collision avoidance procedure in the contention-based phase should be revisited, due to the transitional behavior of interference, to maximize throughput/delay performance of mmWave networks.
We conclude that, unless proper hybrid schemes are investigated, the severity of the transitional behavior may significantly reduce throughput/delay performance of mmWave networks.
\end{abstract}


\section{Introduction}\label{sec: introductions}
Increased demands for extremely high data rates and limited available spectrum for wireless systems in the UHF bands (below 3~GHz) motivate the use of millimeter wave (mmWave) communications to support multi-gigabit data rates. MmWave communication can support many diverse applications including Gbps short range wireless kiosks, augmented reality, massive wireless access in crowd public places, intra- and inter-vehicles connections, wireless connections in data centers, and mobile fronthauling and backhauling. This vast range of applications has led to several standardization activities such as ECMA~387~\cite{ECMA387}, IEEE~802.15.3c~\cite{802_15_3c}, IEEE~802.11ad~\cite{802_11ad}, WirelessHD consortium, wireless gigabit alliance (WiGig), and recently IEEE~802.11ay, established in May~2015.\footnote{Detailed information about these projects can be found at the following addresses: \url{http://www.wirelesshd.org} (WirelessHD), \url{http://wirelessgigabitalliance.org} (WiGig), and \url{http://www.ieee802.org/11/Reports/ng60_update.htm} (802.11ay), respectively.}
The Federal Communications Commission in the USA and the Ofcom in UK also published individual notice of inquiries in early 2015 to investigate if the mmWave bands should be re-purposed for mobile radio services~\cite{FCC2,Ofcom1}. Such evident interests in academia, industry, and regulatory bodies clearly show that mmWave communication technologies will be major components of future wireless networks~\cite{Rappaport2013Millimeter,Andrews2014What,osseiran2014scenarios,boccardi2014Five,shokri2015mmWavecellular,Niu2015Survey}.

MmWave communications use the part of the electromagnetic spectrum between 30 and 300~GHz, which corresponds to wavelengths from 10~mm to 1~mm. The main characteristics of the mmWave communications are high path-loss (distance-dependent component and atmospheric absorption), large bandwidth, short wavelength, and high penetration loss (called blockage)~\cite{Rangan2014Millimeter}. Very small wavelengths allow the implementation of many antenna elements in the current size of radio chips, which promises a substantial increment in the link budget using beamforming. Such a gain can largely or even completely compensate for both the high path-loss and the high noise power (which is due to very large bandwidth) without additional transmission power. Achieving this gain requires having narrow beams both at the transmitter and at the receiver. These narrow-beams, besides boosting the link budget, reduce the interference from other transmitters~\cite{shokri2015mmWavecellular}. In the extreme case, once such multiuser interference is no longer the main limiting factor of the throughput performance, we may face a noise-limited network where the achievable throughput is limited by the noise power.\footnote{Rigorously speaking, negligible multiuser interference does not necessarily imply that the noise power is the main bottleneck of the network throughput performance. Other sources such as beamforming (beam training) overhead may impact the achievable performance of a mmWave network~\cite{Shokri2015mmWaveWPAN}. In this paper, however, we focus on the interference behavior and neglect those overheads, and therefore the communication performance will be limited only by interference and noise powers.} The fundamental question is whether a mmWave network with narrow-beam operation is noise-limited as opposed to the conventional interference-limited networks.
This is a fundamental question at the medium access control (MAC) layer; the answer will reveal the required complexity (and intelligence) that MAC layer functions should support for efficient communications.

The network operating regime may determine which MAC protocol is better suited. For example, the optimal spatial time division multiple access (STDMA) protocol activates a set of transmitter-receiver pairs (links) with negligible mutual interference at a time slot, offering the maximum throughput for every link and for the network~\cite{nelson1985spatial,an2008directional,Shokri2015Beam,bjorklund2003resource}. However, it requires knowledge of precise network topology a priori~\cite{Shokri2015Beam}, which is not available in most of indoor WPAN scenarios, especially those with mobile devices. On the one hand, scheduling based on partial knowledge of the network topology leads to a significant network throughput drop, e.g., 33\% loss is reported in~\cite{gronkvist2001comparison}. On the other hand, discovering the topology (even partial knowledge) requires exchanging several control messages. Sending these control messages may be overwhelming in mmWave networks due to the characteristics of the physical control channel~\cite{shokri2015mmWavecellular}.\footnote{Due to high reliability and robustness requirements, the physical control channel has a significantly lower transmission rate compared to the data channel. IEEE~802.11ad, for instance, supports up to 27.7~Mbps for control packets (a ``packet'' is a message frame at the MAC layer) while 6.7~Gbps is supported for data packets~\cite{Nitsche2014IEEE}. Moreover, sending control packets in the mmWave bands may impose additional beam training overhead compared to sending those in the UHF bands~\cite{shokri2015mmWavecellular}. This alignment is necessary to avoid \emph{deafness}, formally defined later in this section.}
Moreover, the optimal STDMA needs to solve an NP-hard problem for a given network topology~\cite{ramanathan1997unified,gronkvist2001comparison,bjorklund2003resource}, which may lead to largely suboptimal solutions in a network with very fast rescheduling requirements such as in mmWave networks~\cite{shokri2015mmWavecellular}.
To mitigate unaffordable signaling and computational overhead of STMDA, current mmWave standards adopt a very conservative approach of activating only one link at a time through a time division multiple access (TDMA)-based resource allocation~\cite{802_15_3c,802_11ad}. This conservative resource allocation, once again, is substantially suboptimal in mmWave networks~\cite{rhee2008z,son2012frame,Shokri2015Beam,niu2015exploiting}, though achieves the performance of STDMA if there is strong interference between any pair of links. The latter is very unlikely in mmWave networks with narrow-beam operation. Slotted ALOHA, as an alternative contention-based resource allocation solution, imposes no signaling and computational overhead and achieves the performance of STDMA provided that there is no mutual interference between any pair of links (a noise-limited regime). Simple protocols such as carrier sense multiple access (CSMA) and CSMA with collision avoidance (CSMA/CA) are the most common modifications of slotted ALOHA to regulate multiple access without network wide synchronization or global topology information. CSMA/CA is substantially throughput-suboptimal due to the overhead of collision avoidance messages~\cite{magistretti2014802}, yet it alleviates hidden and exposed node problems, and thereby can outperform CSMA. However, all these contention-based protocols cannot guarantee collision-free communications, which is important in many applications. Hybrid MAC approaches, mainly developed for interference-limited networks, can combine the strengths and offset the weaknesses of contention-based and contention-free resource allocation strategies~\cite{ephremides1982analysis,rios1985hybrid,van2003adaptive,ye2004medium,vuran2007mac,rhee2008z}.

To design a proper hybrid MAC for mmWave networks with narrow-beam operation, the first steps are analyzing the collision, evaluating performance gain (in terms of throughput/delay) due to various resource allocation protocols, and investigating the signaling and computational complexities of those protocols. Roughly speaking, as the system goes to the noise-limited regime, the required complexity for proper resource allocation and interference avoidance functions at the MAC layer substantially reduces~\cite{Singh2011Interference,Qiao2015D2D,Niu2015Blockage,Son2012on,Park2009Analysis,Shokri2015Beam}. For instance, in a noise-limited regime, a very simple resource allocation such as activating all links at the same time without any coordination among different links may outperform a complicated independent-set based resource allocation~\cite{Shokri2015Beam}. Instead, narrow-beam operation complicates negotiation among different devices in a network, as control message exchange may require time consuming antenna alignment procedure to avoid \emph{deafness}~\cite{Shokri2015Beam}. Deafness refers to the situation in which the main beams of the transmitter and the receiver do not point to each other, preventing establishment of a communication link.
Therefore, determining the network operating regime is essential to determine the best MAC layer protocol. How to make such a determination is largely an open problem for mmWave networks.

The seminal work in~\cite{Singh2011Interference} shows the feasibility of \emph{pseudo-wired} abstraction (noise-limited network) in outdoor mmWave mesh networks. However, as shown in~\cite{Qiao2012STDMA,Sum2009virtual,Park2009Analysis,Shokri2015Beam}, indoor mmWave WPANs are not necessarily noise-limited. In particular, activating all links causes a significant performance drop compared to the optimal resource allocation~\cite{Shokri2015Beam}, indicating that there may be situations in which a non-negligible multiuser interference is present; the noise power is not always the limiting factor.
Such a performance degradation increases with the number of devices in the network~\cite{Shokri2015Beam}. This indeed means that the accuracy of the noise-limited assumption to model the actual network behavior reduces with the number of links. Similar conclusions are also made in mmWave cellular networks~\cite{di2014stochastic}.
It follows that adopting the noise-limited assumption may be detrimental for proper MAC layer design. However, the interference footprint may not be so large that we need to adopt very conservative resource allocation protocols such as TDMA, which activates only one link at a time.

In this paper, we investigate the fundamental performance indicators that will help in deciding which MAC is the best for which situation. To this end, we first introduce a novel blockage model that, unlike the existing models~\cite{di2014stochastic,park2015TractableResource,lu2015stochastic,Singh2015TractableModel,TBai2014Coverage}, captures the angular correlation of the blockage events as a function of size and density of the obstacles. We drive tractable closed-form expressions for collision probability, per-link throughput, and area spectral efficiency. We analytically evaluate the impact of the transmission/reception beamwidth, transmission power, and the densities of the transmitters and obstacles on the performance metrics. The new analysis shows that the pseudo-wired abstraction may not be accurate even for a modest-sized ad~hoc network, and mmWave networks exhibit a transitional behavior from a noise-limited regime to an interference-limited regime. Using the established collision analysis, we investigate if either a contention-based or contention-free resource allocation protocol is a good option for a mmWave ad~hoc network. To this end, we derive the exact expressions and tight bounds on the MAC layer throughput of a link, area spectral efficiency, and delay performance of STDMA, TDMA, and slotted ALOHA protocols. We also numerically evaluate those metrics for CSMA and CSMA/CA. Comprehensive analysis reveals that STDMA is impractical due to massive signaling and computational overheads. Conventional CSMA/CA is very throughput/delay inefficient due to unnecessary overhead of the collision avoidance procedure. A simple CSMA (or even slotted ALOHA) may achieve the performance of STDMA and may significantly outperform TDMA in terms of network throughput/delay performance, whereas TDMA is still necessary to guarantee collision-free communications. We conclude that the transitional behavior of interference in mmWave networks necessitates a collision-aware hybrid resource allocation procedure, containing both contention-based and contention-free phases with flexible phase duration. In particular, the contention-based phase with on-demand execution of the collision avoidance function substantially improves throughput/delay performance of the network. Moreover, on-demand use of the contention-free phase to deliver only the collided packets guarantees a reliable physical layer with minimal drop in the network throughput/delay performance.
Detailed analysis of this paper clarifies the collision level and throughput performance of mmWave networks, and thereby provides useful guidelines for designing proper resource allocation and interference management protocols for future mmWave networks.

The rest of this paper is organized as follows. In Section~\ref{sec: system model}, we describe the system model. The collision probability in mmWave ad~hoc networks is derived in Section~\ref{sec: collision-analysis}, followed by evaluation of the MAC throughput and characterization of the network operating regime in Section~\ref{sec: Performance-analysis}. The paper is concluded in Section~\ref{sec: Conclusion}.

\section{System Model}\label{sec: system model}
\nocite{Singh2011Interference,TBai2014Coverage,hunter2008transmission,wildman2014joint,shokri2015MillimeterGlobecom,pyo2009throughput}
We consider a mmWave wireless network, and a homogeneous Poisson network of transmitters on the plane with density $\lambda_t$ per unit area, each associated to a receiver. To evaluate the collision performance of the network, we consider a reference link (called typical link) between a reference receiver and its intended transmitter having geometrical/spatial length $L$, see Table~\ref{table: notations} for a list of the main symbols used in the paper.
We call the receiver and the transmitter of the typical link as the typical receiver and the tagged transmitter.
From Slivnyak's Theorem~\cite[Theorem 8.1]{Haenggi2013Stochastic} applied to homogeneous Poisson point processes, the conditional distribution of the potential interferers (all transmitters excluding the tagged transmitter) given the typical receiver at the origin is another homogeneous Poisson point process with the same density. We assume that if multiple neighbors are transmitting to the same receiver, at most one of them can be successfully decoded by that receiver. This natural assumption, as commonly considered in the performance evaluation~\cite{pyo2009throughput,Singh2011Interference,TBai2014Coverage,hunter2008transmission,wildman2014joint,shokri2015MillimeterGlobecom}, is motivated by the lack of multiuser detection in many devices including mmWave ones~\cite{802_15_3c,802_11ad}.
Therefore, all transmitters in the network act as potential interferers for the typical receiver (the receiver of the typical link).
The interference level depends on the density and location of the interferers relative to the typical receiver, transmission powers, channel model, antenna radiation pattern, blockage model, and transmission and reception beamwidths.

\begin{table}[t]
  \centering
  \caption{Summary of main notations}\label{table: notations}
{
\renewcommand{\arraystretch}{1.2}
  {
   \begin{tabular}{|@{}c@{}|l|}
\hline
   \textbf{Symbol} & \textbf{Definition} \\ \hline
    $A_d$ & Area of circle sector with radius $d$ and angle $\theta_c$ \\ \hline
    ${\mathrm{ASE}}_{\text{S-ALOHA}}$ & Area spectral efficiency of slotted ALOHA \\ \hline
    ${\mathrm{ASE}}_{\text{TDMA}}$ & Area spectral efficiency of TDMA \\ \hline
    $d_{\max}$ & Interference range \\ \hline
    $L$ & Geographical/spatial length of the typical link \\ \hline
    $n_I$ & The number of interferers \\ \hline
    $n_o$ & The number of obstacles \\ \hline
    $r_{_{\text{S-ALOHA}}}$ & Average throughput of a link in slotted ALOHA \\ \hline
    $r_{_{\text{TDMA}}}$ & Average throughput of a link in TDMA \\ \hline
    $\theta$ & Transmission/reception beamwidth \\ \hline
    $\theta_c$ & Coherence angle \\ \hline
    $\lambda_I$ & Density of potential interferers per unit area\\ \hline
    $\lambda_t$ & Density of transmitters (links) per unit area \\ \hline
    $\lambda_o$ & Density of obstacles per unit area\\ \hline
    $\rho_a$ & Transmission probability of slotted ALOHA \\ \hline
    $\rho_{c \mid L} \left(\ell \right)$ & Conditional collision probability given $L$ \\ \hline
    $\rho_{s \mid L} \left(\ell \right)$ & Conditional probability of successful transmission given $L$ \\ \hline
\end{tabular}}
}
\end{table}

We consider a slotted ALOHA protocol without power control to derive a lower bound on the performance.\footnote{Kleinrock's seminal work shows that simple CSMA protocols easily outperform both pure and slotted ALOHA protocols~\cite{kleinrock1975packet}. As will be shown in this paper, there is a non-negligible contention on the channel access, making it imperative to add a simple carrier sense functionality to the slotted ALOHA. However, as the system goes to the noise-limited regime, the performance gain due to this additional functionality vanishes.} That is, the transmission power of all links is $p$. We let every transmitter (interferer) be active with probability $\rho_a$, so the probability of transmitting in a slot is $\rho_a$. In the slotted ALOHA, the transmissions are regulated to start at the beginning of a time slot. The slotted ALOHA is a good model for the worst case analysis of a device-to-device (D2D) network underlaying a cellular network, as devices are synchronous by using base station synchronization signals. Also, slotted ALOHA provides an upper bound on the throughput performance of pure ALOHA, where the transmission is started immediately upon a new packet arrival~\cite{kleinrock1975packet}. Although for analytical tractability we choose slotted ALOHA, the analysis of this paper can be readily extended to the pure ALOHA case. Further, similar to~\cite{hunter2008transmission,wildman2014joint}, we assume that transmitter of every link is spatially aligned with its intended receiver, so there is no beam training overhead.
The adverse impacts of the beam training overhead on per-link and network throughput performance are investigated in~\cite{Shokri2015Beam}. In this paper, instead, we have assumed pre-aligned transmitter-receiver pairs to analyze the impact of other parameters (such as density of the transmitters, operating beamwidth, density and size of the obstacles, and the blockage model) on the performance of mmWave networks. Note that the beam training procedure imposes the same overhead on all resource allocation protocols we are considering in this paper, so it can be neglected from the comparative analysis and conclusions. If there is no obstacle on the link between transmitter $i$ and the origin, we say that transmitter $i$ has LoS condition with respect to the typical receiver, otherwise it is in non-LoS condition. Moreover, similar to~\cite{Singh2011Interference,hunter2008transmission,wildman2014joint,pyo2009throughput,TBai2014Coverage,di2014stochastic,Singh2015TractableModel,park2015TractableResource,lu2015stochastic}, we consider only LoS links and neglect reflections in mathematical analysis, but discuss its impacts on both collision probability and throughput.

We consider a distance-dependent path-loss with exponent $\alpha$, as commonly assumed for MAC layer performance evaluations~\cite{singh2009blockage,Singh2011Interference}.
This simple model allows deriving tractable closed-form expressions for the collision probability and for the throughput, and, at the same time, enables us to draw general conclusions about the network operating regime.
Note that the sparse scattering feature of mmWave frequencies, along with pencil-beam operation, makes the mmWave channel more deterministic compared to that of the conventional systems, which normally operate in rich scattering environments and with omnidirectional communication~\cite{Rappaport2015wideband}. Moreover, in the mathematical analysis, we have ignored extra attenuation due to the atmospheric absorption at the mmWave frequencies. This is motivated by negligible extra channel attenuation ($<$~0.3~dB) for typical ranges of the mmWave networks, i.e., less than 300~m for cellular networks (e.g., at 28~GHz) and less than 15~m for short range networks (e.g., at 60~GHz)~\cite{Rappaport2015wideband}. For instance, this extra channel attenuation is around 0.05~dB at 28~GHz with the atmospheric absorption of 0.15~dB/Km, and it is around 0.24~dB at 60~GHz with 16~dB/Km atmospheric absorption~\cite{Rangan2014Millimeter}. However, we include its effects in the numerical analysis.

We use the \emph{protocol model} of interference~\cite{gupta2000capacity}, as it is common for the MAC layer analysis~\cite{Singh2011Interference,xu2002effective,iyer2009right,cardieri2010modeling}. In this model, for a given distance between a reference receiver and its intended transmitter, a \emph{collision}\footnote{Note that ``collision'' is defined as the outage event due to strong interference from other transmitters. Note that an outage can also occur due to low signal-to-noise ratio (SNR) even without any interference.} occurs if there is at least another interfering transmitter no farther than a certain distance from the reference receiver, hereafter called \emph{interference range}.
Besides its simplicity, our recent investigation in~\cite{Shokri2015OntheAccuracy} reveals that the special characteristics of mmWave networks makes such interference model quite accurate for them. Essentially, as the probability of having LoS condition on a link decreases exponentially with the distance~\cite[Fig.~4]{Rappaport2015wideband}, far away transmitters will be most probably blocked (in non-LoS condition) and therefore cannot contribute in the interference a receiver experiences. Therefore, we may consider only the impact of spatially close transmitters, and yet have negligible loss in the accuracy of the interference model. Moreover, due to directional communications, a small number of those close transmitters can cause non-negligible interference at the receiver side, further increasing the accuracy of the protocol model, see~\cite{Shokri2015OntheAccuracy} for detailed discussions.

At the MAC layer, the beamforming is represented by using an ideal sector antenna pattern~\cite{hunter2008transmission,wildman2014joint,TBai2014Coverage}, where the directivity gain is a constant for all angles in the main lobe and equal to a smaller constant in the side lobe.
This model allows capturing the interplay between antenna gain, which ultimately affects the transmission range, and the half power beamwidth. We assume all devices in both transmission and reception modes operate with the same beamwidth $\theta$. Considering 2D beamforming, the directivity gain for each transmitter/receiver is
\begin{equation}\label{eq: antenna-pattern}
g = \left\{{\begin{array}{*{20}{l}}
\frac{2 \pi - (2\pi - \theta)\epsilon}{\theta} \:, & {\text{in the main lobe}}\\
\epsilon \:, & {\text{in the side lobe}}
\end{array}} \right. \:,
\end{equation}
where typically $0 \leq \epsilon \ll 1$. The gain in the main lobe can be derived by fixing the total radiated power of the antennas over parameter space of $\epsilon$ and $\theta$. Due to small value of $\epsilon$ compared to the directivity gain in the main lobe, only the interferers that are aligned with the typical receiver can cause collision. In other words, there is no strong interference, so no collision, in the deafness condition. Detailed quantitative analysis of~\cite{Shokri2015WhatIs} shows that neglecting side lobe transmissions from the interference model is valid for a system with more than 15~dB side lobe suppression, which is easy to achieve in the mmWave systems~\cite{Rangan2014Millimeter}.

Further, the extremely high penetration loss in mmWave networks almost vanishes the impact of any transmitter with non-LoS condition with respect to a receiver. To have quantitative insights, mmWave signals will be attenuated by 20-35~dB due to the human body~\cite{Rangan2014Millimeter}. This extreme penetration loss not only blocks a link between a receiver and its intended transmitter, as argued in~\cite{singh2009blockage}, it also vanishes the impact of unintended transmitters with non-LoS conditions (non-LoS interferers) on the aggregated interference level the receiver experiences. The negligible impact of the non-LoS interferers is also confirmed in~\cite{Shokri2015WhatIs}.


Due to sensitivity of the mmWave links to any obstacle, the first step in analyzing the system-level performance of mmWave networks is introducing a blockage model. A proper blockage model should capture the following properties: (\textit{i}) obstacles may randomly appear in a communication link and (\textit{ii}) one obstacle may block multiple angularly close communication links (angular correlation). Using the random shape theory,~\cite{TBai2014Blockage} proposes a simple blockage model for urban mmWave cellular networks that addresses property (\textit{i}). In this model, the event of having obstacles in the link between any transmitter-receiver is independent of all other links and increases exponentially with the link length. This model is approximated by a LoS ball model~\cite{TBai2014Coverage}, wherein all transmitters within a certain distance of any receiver (inside a ball centered at the location of that receiver) observe the LoS condition, and all other transmitters outside the ball observe the non-LoS condition with respect to the reference receiver.~\cite{Singh2015TractableModel} augments the LoS ball model by a Bernoulli process, i.e., each transmitter inside the ball is in the non-LoS condition with a constant (non-zero) probability, still outside transmitters are always in the non-LoS condition.~\cite{di2014stochastic} extends this model to a two-ball model, in which the transmitters located outside the outer ball are always in outage.~\cite{Akdeniz2014MillimeterWave} models the blockage with a random attenuation with a given density, whose parameters are derived from the channel measurements. Though being used for performance evaluation, all these blockage models share the same drawback: they fail to capture angular correlation of the LoS events. As the operating beamwidth becomes narrower, the events of observing obstacles on the link between a receiver and individual interferers have an increased correlation, so the LoS condition for different interferers becomes correlated. Many interferers that are angularly closely located from the point of view of the receiver can be blocked by an obstacle between them and the receiver. The accuracy of the assumption of independent LoS conditions on the links among the typical receiver and different interferers decreases either if we increase the density of the transmitters or if the transmitters appear in spatial clusters. The consequence is that those blockage models may sometimes prevent deriving correct conclusions, especially for dense mmWave networks.

\noindent \textbf{Blockage model:} In this paper, assuming that the centers of the obstacles\footnote{For sake of simplicity, we may use obstacle to refer the center of that obstacle throughout the paper.} follow a homogeneous Poisson point process with density $\lambda_o$ independent of the communication network, we use the following model to capture the aforementioned angular correlation among LoS conditions: we define a \emph{coherence angle} $\theta_c$ over which the LoS conditions are statistically correlated. That is, inside a coherence angle, an obstacle blocks all the interferers behind itself, so there is no LoS conditions in distances $d \geq l$ with respect to the receiver of the typical link and consequently no LoS interferers, if there is an obstacle at distance $l$. However, there is no correlation between LoS condition events in different coherence angle intervals, i.e., in different circle sectors with angle $\theta_c$. The coherence angle increases with the size and density of the obstacles in the environment. In this paper, we assume that $\theta_c$ is constant and given. Exact characterization of the coherence angle as a function of the size and density of the obstacles and interferers is the subject of our future studies. Note that different obstacles with different sizes and locations can cause different intervals $\theta_c$ of the angular correlation of blockage events. However, we suggest using the average value of $\theta_c$ to simplify the analysis, which otherwise would be intractable. We made this proposal inspired by the classic concepts of the coherence time and the coherence bandwidth for wireless channels. The coherence time and coherence bandwidth are different for different users with different speeds and different surrounding environments; still, the common approach is assuming the same values for all users to simplify the analysis (see~\cite{goldsmith2003capacity} and references therein).
Using the average coherence angle in the proposed blockage model indeed imply that this model is suitable for \emph{ergodic} system-level performance analysis, where the achieved performance metrics are averaged over sufficiently large number of realizations of the obstacle process. In other words, to derive ergodic performance metrics, we can consider the proposed blockage model to well approximate the individual realizations of the actual blockage process.

For mathematical tractability, we need the following main assumptions: \textit{i}) protocol model of interference, \textit{ii}) constant coherence angle for all realizations of the obstacle process with a given average size and density of the obstacles, and \textit{iii}) independent number of LoS interferers in different coherent angle intervals.
With these simplifying assumptions, in the following, we derive closed-form expressions for the collision probability, per-link throughput, and area spectral efficiency. Then, we show a well coincidence between the derived equations (which include these simplifying assumptions) with the reality (which does not have those assumption), validating those simplifying assumptions.

\section{Collision Analysis}\label{sec: collision-analysis}
In this section, we investigate the collision probability in a mmWave network working with slotted ALOHA protocol. The derivation of such a result will play a major role in performance analysis of mmWave networks, presented in Section~\ref{sec: Performance-analysis}.

We consider a typical receiver at the origin of the Polar coordinates and its intended transmitter at distance $L$ and evaluate the collision probability due to other transmitters' operation located inside the circle sector with angle $\theta$ and radius of the interference range.
Let $p$ be the transmission power, and $a$ be the average channel attenuation at reference distance 1~meter. The channel gain between the typical receiver and an aligned non-blocked transmitter at distance $d$ is $a d^{-\alpha}$.

We denote by $d_{\max}$ the interference range, by $\beta$ the minimum SINR threshold at the typical receiver, and by $\sigma$ the noise power. The interference range $d_{\max}$ is defined as the maximum distance an interferer can be from the receiver and still cause collision/outage.
At the typical receiver, the SINR due to transmission of the intended transmitter and an aligned LoS interferer located at distance $d$ is:
\begin{equation*}
\mathrm{SINR} = \frac{p \left( \frac{2 \pi - (2\pi - \theta)\epsilon}{\theta} \right)^2 a {L}^{-\alpha} }{p \left( \frac{2 \pi - (2\pi - \theta)\epsilon}{\theta} \right)^2 a d^{-\alpha} +\sigma} \:.
\end{equation*}
Comparing the SINR expression to $\beta$, we get the interference range
\begin{equation}\label{eq: d_max}
d_{\max} = \left( \frac{L^{-\alpha}}{\beta} - \frac{\sigma}{pa}\left( \frac{\theta}{2 \pi - (2\pi - \theta)\epsilon}\right)^{2} \right)^{{-1/\alpha}} .
\end{equation}
Note that changing the channel model affects only $d_{\max}$ and all the following expressions will be valid by substituting the new $d_{\max}$. For instance, to consider 60~GHz communications and introduce the exponential atmospheric absorption (16~dB/Km extra attenuation~\cite{Rangan2014Millimeter}) into the analysis, we only need to change the channel model from $ad^{-\alpha}$ to $ad^{-\alpha}e^{-0.0037d}$ and find $d_{\max}$ from the new SINR expression, see~\cite[Equations~(1) and (9)]{Singh2011Interference}.

\begin{figure}[!t]
\centering
  \includegraphics[width=0.85\columnwidth]{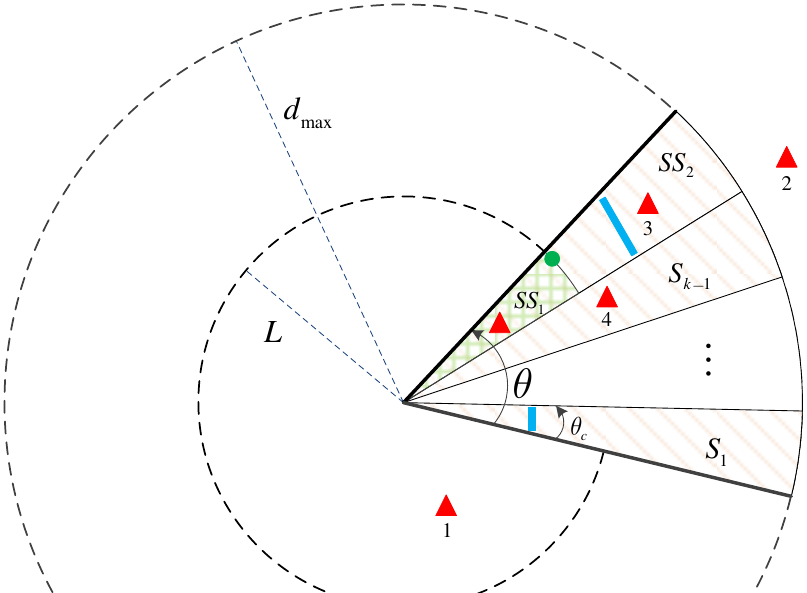}\\

  \caption{Hatched lines show potential interference zone. Operating beamwidth $\theta$ is divided into $k$ sectors of angle $\theta_c$. The typical receiver is on the origin. The tagged transmitter, shown by a green circle, is on sector $k$ at distance $L$ of the typical receiver. $S_i$ shows sector $1 \leq i \leq k-1$. ${SS}_1$ and ${SS}_2$ are two sub-sectors of sector $k$. Zones with orange hatched lines have both random interferers and obstacles, represented by a red triangle and a blue rectangle. Zones with green hatched lines have only random interferers. $d_{\max}$ is the interference range.}
  \label{fig: IntRegion}
\end{figure}

A transmitter at distance $d$ from the typical receiver can cause collision provided that the following conditions hold: (a) it is active, (b) the typical receiver is inside its main lobe, (c) it is inside the main lobe of the typical receiver, (d) it is located inside the interference range $d \leq d_{\max}$, and (e) it is in the LoS condition with respect to the typical receiver. These conditions are illustrated in Fig.~\ref{fig: IntRegion}, where the tagged transmitter, interferers, and obstacles are represented by a green circle, red triangles, and blue rectangles, respectively.
Also, the highlighted part is the sector from which the typical receiver is receiving signal.
Interferers~1, 2, and 3 cannot cause collision at the typical receiver due to condition~(c),~(d), and~(e), respectively.
Due to random deployment of the devices, the probability that the typical receiver locates inside the main lobe of an active transmitter is $\theta/2 \pi$. Therefore, if the density of transmitters per unit area is $\lambda_t$ and if the average probability of being active for every transmitter is $\rho_a$, the interferers for which conditions~(a) and~(b) hold follow a homogeneous Poisson point process with density $\lambda_I = \rho_a \lambda_t \theta/ 2 \pi$ per unit area. Conditions~(c) and~(d) reduces the area over which a potential interferer can cause collision. For condition~(e), we need to elaborate the blockage model. The typical receiver observes $k = \lceil \theta / \theta_c \rceil $ sectors, each with angle $\theta_c$, where $\lceil \cdot \rceil$ is the ceiling function. For the sake of simplicity, we assume that $ \theta / \theta_c $ is an integer; however the analysis can be extended, with more involved calculations, to the general case. We take the general assumption that the tagged transmitter is uniformly distributed in the circle sector with angle $\theta$ that the typical receiver is pointing to, as shown by hashed lines in Fig.~\ref{fig: IntRegion}. Having a fix coordinate for the tagged transmitter is a special case of our analysis. It is straightforward to see that the tagged transmitter is located in one of these $k$ sectors with uniform distribution and its radial distance to the typical receiver $L$ is a continuous random variable with density function $f_L(\ell) = 2\ell / d_{\max}^{2}$. Without loss of generality, we assume that the tagged transmitter is in sector $k$. It means that we have a combination of interferers and obstacles in the first $k-1$ sectors. In the last sector, we cannot have any obstacle in the circle sector with angle $\theta_c$ and radius $L$, as the tagged transmitter in $L$ should be in the LoS condition, otherwise the typical link will not be established and collision cannot happen. Dividing the last sector into two sub-sectors, corresponding to the distances $\left( 0, L \right]$ and $\left( L , d_{\max}\right]$, the first sub-sector contains only interferers, whereas the second one has both interferers and obstacles.
In the following, we first derive the probability of receiving collision from individual sectors and then compute the collision probability in general.

Let $A_d$ be the area of a circle sector with radius $d$ and angle $\theta_c$. The number of interferers and obstacles in every sector $s$, $1 \leq s \leq k-1$, respectively denoted by $n_I$ and $n_o$, are independent Poisson random variables with average $\lambda_I A_d$ and $\lambda_o A_d$. Given sector $s$, $1 \leq s \leq k-1$, we have three possible cases:
\begin{enumerate}
  \item $n_I = 0  , n_o \geq 0$: There is no interferer, and consequently the probability of LoS interference is 0.
  \item $n_I \geq 1  , n_o = 0$: In this case, every interferer in the sector is a LoS interferer that causes collisions. The probability of LoS interference in this case is 1.
  \item $n_I \geq 1  , n_o \geq 1$: In this case, we have a combination of interferes and obstacles located randomly inside the sector.
  Let $\{X_{1},X_{2},\ldots,X_{n_I}\}$ and $\{Y_{1},Y_{2},\ldots,Y_{n_o}\}$ be the set of distances of $n_I$ interferers and $n_o$ obstacles from the origin. We define random variables ${X_{(1)} = \min \{X_{1},X_{2},\ldots,X_{n_I}\}}$ and ${Y_{(1)} = \min \{Y_{1},Y_{2},\ldots,Y_{n_o}\}}$. Given $n_I \geq 1$ and $n_o \geq 1$, the typical receiver observes at least one LoS interferer provided that $X_{(1)} < Y_{(1)}$. We characterize the probability of having at least one LoS interferer in the following propositions.
\end{enumerate}

\begin{lemma}\label{prop: jointPDF}
Consider the blockage model, described in Section~\ref{sec: system model} and in Fig.~\ref{fig: IntRegion}. Given sector $s$, the number of interferers $n_I \geq 1$, and the number of obstacles $n_o \geq 1$, joint probability density function of $X_{(1)}$, $Y_{(1)}$, $n_I$, and $n_o$ is given by Equation~\eqref{eq: jointPDFprop1} on the top of page~\pageref{eq: jointPDFprop1}. Also, the probability of having at least one LoS interferer given $n_I \geq 1$ and $n_o \geq 1$, denoted by $\Pr [\mathrm{LI} \mid n_I \geq 1, n_o \geq 1]$, is given by Equation~\eqref{eq: LoSIwithJointPDF}.
\begin{figure*}[!t]
\normalsize
\begin{align}\label{eq: jointPDFprop1}
f_{X_{(1)},Y_{(1)},n_I,n_o} \left( x , y, n, m | n \geq 1, m \geq 1 \right)  = & \:
\displaystyle \frac{2nx}{d_{\max}^{2}}\left( 1 - \frac{x^2}{d_{\max}^{2}} \right)^{n-1} \frac{2my}{d_{\max}^{2}} \left( 1 - \frac{y^2}{d_{\max}^{2}} \right)^{m-1} \frac{e^{-\lambda_I A_{d_{\max}}}}{1 - e^{-\lambda_I A_{d_{\max}}}}\frac{\left( \lambda_{I} A_{d_{\max}} \right)^{n}}{n!} \\
\displaystyle & \times
\frac{e^{-\lambda_o A_{d_{\max}}}}{1 - e^{-\lambda_o A_{d_{\max}}}}\frac{\left( \lambda_{o} A_{d_{\max}} \right)^{m}}{m!}  \nonumber
\end{align}
\hrulefill
\begin{equation}\label{eq: LoSIwithJointPDF}
\Pr [\mathrm{LI} \mid n_I \geq 1, n_o \geq 1] = \frac{\lambda_o}{\left( 1 - e^{-\lambda_I A_{d_{\max}}} \right)\left( 1 - e^{-\lambda_o A_{d_{\max}}} \right)}
\left( \frac{1 - e^{-\lambda_o A_{d_{\max}}}}{\lambda_o} -
\frac{1 - e^{-\left( \lambda_o + \lambda_I \right) A_{d_{\max}}}}{\lambda_o + \lambda_I}  \right) \:.
\end{equation}
\hrulefill
\end{figure*}
\end{lemma}

\begin{IEEEproof}
A proof is given in Appendix~A.
\end{IEEEproof}

Using Lemma~\ref{prop: jointPDF}, we can find the probability of having LoS interference in sector $s$, $1 \leq s \leq k-1$.

\begin{prop}\label{prop: LoSI}
Consider the blockage model, described in Section~\ref{sec: system model} and in Fig.~\ref{fig: IntRegion}. Given sector $s$, $1 \leq s \leq k-1$, the probability of having at least one LoS interferer is given by Equation~\eqref{eq: LoSISec1:K-1}, where $\lambda_I = \rho_a \lambda_t \theta / 2 \pi$ and $A_{d_{\max}} = \theta_c d_{\max}^{2}/2$.
\begin{figure*}[!t]
\normalsize
\begin{equation}\label{eq: LoSISec1:K-1}
\Pr[{\text{LoS interference from sector $s$, $1 \leq s \leq k-1$}}] = \frac{\lambda_I}{\lambda_o + \lambda_I} \left( 1 - e^{-\left( \lambda_o + \lambda_I \right) A_{d_{\max}}} \right) \:.
\end{equation}
\hrulefill
\end{figure*}
\end{prop}

\begin{IEEEproof}
For sake of notation simplicity, we denote by $\Pr[{\mathrm{LI}}]$ the probability of having at least one LoS interferer in a given sector $s$, $1 \leq s \leq k-1$. Let $n_I = n$ and $n_o = m$. Considering the discussions at the beginning of this subsection and mutual independence of the number of interferes and obstacles, we have~\eqref{eq: LIprob}, where $\Pr[{\mathrm{LI}} ~| n \geq 1, m = 0] = 1$, $\Pr[ n \geq 1] = 1 - e^{-\lambda_I A_{d_{\max}}}$, $\Pr[ m = 0] = e^{-\lambda_o A_{d_{\max}}}$, ${\Pr[{\mathrm{LI}} ~| n \geq 1, m \geq 1]}$ is given in~\eqref{eq: LoSIwithJointPDF}, and $\Pr[ m \geq 1] = 1 - e^{-\lambda_o A_{d_{\max}}}$.
After some algebraic manipulations, we have~\eqref{eq: LIfinal}, which concludes the proof.
\begin{figure*}[!t]
\normalsize
\begin{align}\label{eq: LIprob}
\Pr[{\mathrm{LI}}] &= \Pr[{\mathrm{LI}} ~| n = 0]\Pr[ n = 0] + \Pr[{\mathrm{LI}} ~| n \geq 1, m = 0] \Pr[ n \geq 1, m = 0] +\Pr[{\mathrm{LI}} ~| n \geq 1, m \geq 1] \Pr[ n \geq 1, m \geq 1]  \\
&= \Pr[{\mathrm{LI}} ~| n \geq 1, m = 0] \Pr[ n \geq 1]\Pr[m = 0] +
\Pr[{\mathrm{LI}} ~| n \geq 1, m \geq 1] \Pr[ n \geq 1]\Pr[m \geq 1]  \:. \nonumber
\end{align}
\hrulefill
\begin{equation}\label{eq: LIfinal}
\Pr[{\mathrm{LI}}] =
\left(1 - e^{-\lambda_I A_{d_{\max}}} \right) e^{-\lambda_o A_{d_{\max}}} + \lambda_o \left( \frac{1 - e^{-\lambda_o A_{d_{\max}}}}{\lambda_o} -
\frac{1 - e^{-\left( \lambda_o + \lambda_I \right) A_{d_{\max}}}}{\lambda_o + \lambda_I}  \right)
=\frac{\lambda_I}{\lambda_o + \lambda_I} \left( 1 - e^{-\left( \lambda_o + \lambda_I \right) A_{d_{\max}}} \right) \:.
\end{equation}
\hrulefill
\vspace*{4pt}
\end{figure*}
\end{IEEEproof}

In order to numerically illustrate Proposition~\ref{prop: LoSI} and derive some insights on the behavior of LoS interference probability formulated in~\eqref{eq: LoSISec1:K-1}, we simulate an ad~hoc network with random number of mmWave links, operating with beamwidth $\theta = 20 \degree$ at 60~GHz. The transmission probability of every link is 1, so all links are always active. We assume 2.5~mW transmission power, 16~dB/Km atmospheric absorption, coherence angle $\theta_c = 5 \degree$, and interference range $d_{\max} = 15$~m. Using Monte Carlo simulations, we evaluate the average probability of having a LoS interference over $10^6$ random topologies.
Changing $\lambda_t$, $\lambda_o$, $\theta$, and $d_{\max}$ we can cover a wide variety of future mmWave applications, including:
\begin{itemize}
  \item long range, low mobility, low density applications such as mobile fronthauling and backhauling use cases, which correspond to high $d_{\max}$ and small $\theta$, $\lambda_o$, and $\lambda_t$; and
  \item short range, high mobility, massive wireless access applications such as crowded public place use case, which correspond to small $d_{\max}$, relatively wide $\theta$, and high $\lambda_o$ and $\lambda_t$.
\end{itemize}

\begin{figure}[!t]
	\centering
    \subfigure[]{
	\includegraphics[width=\columnwidth]{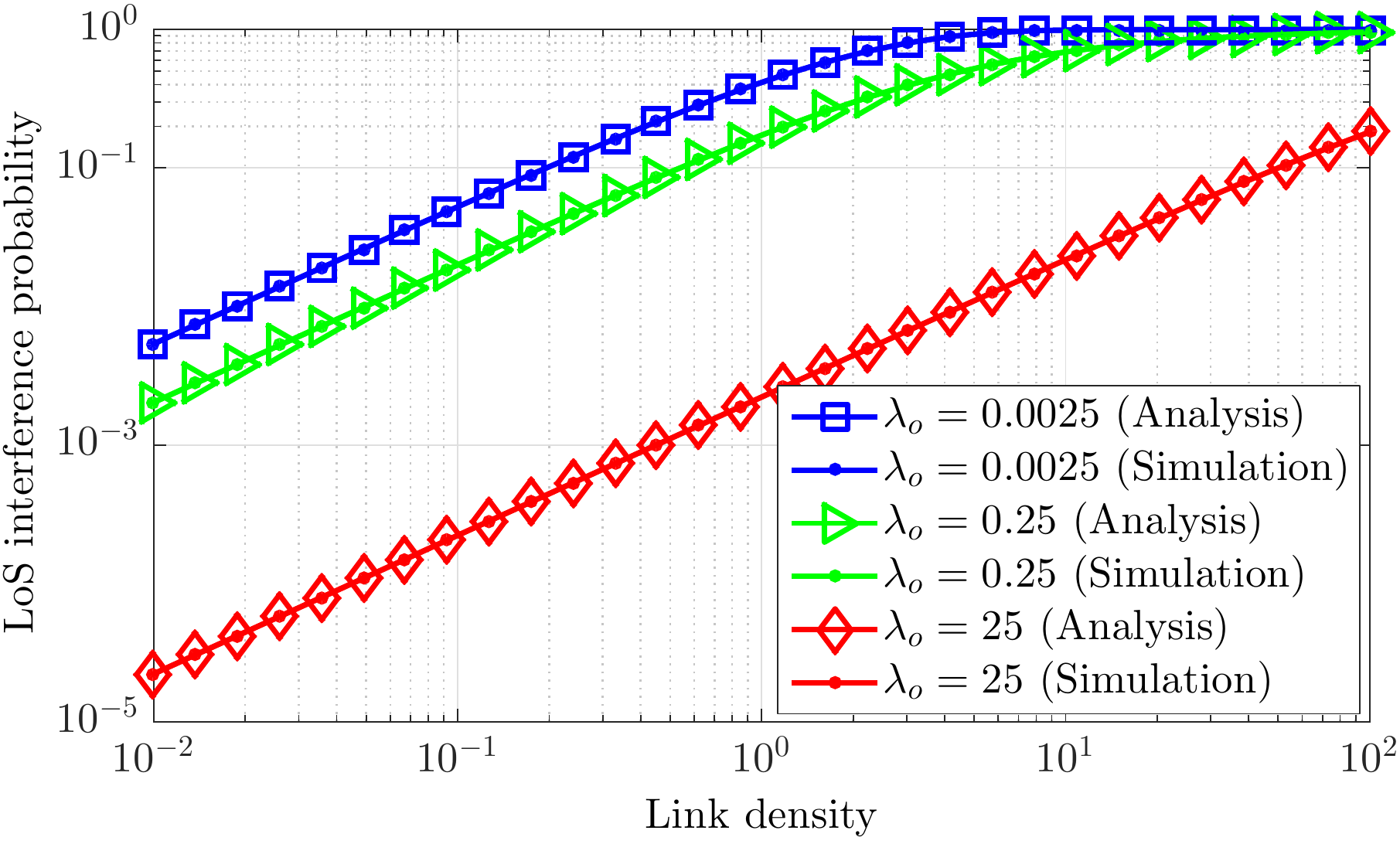}
		\label{subfig: LoSInt_SingleSector_I}
	}
	\subfigure[]{
	\includegraphics[width=\columnwidth]{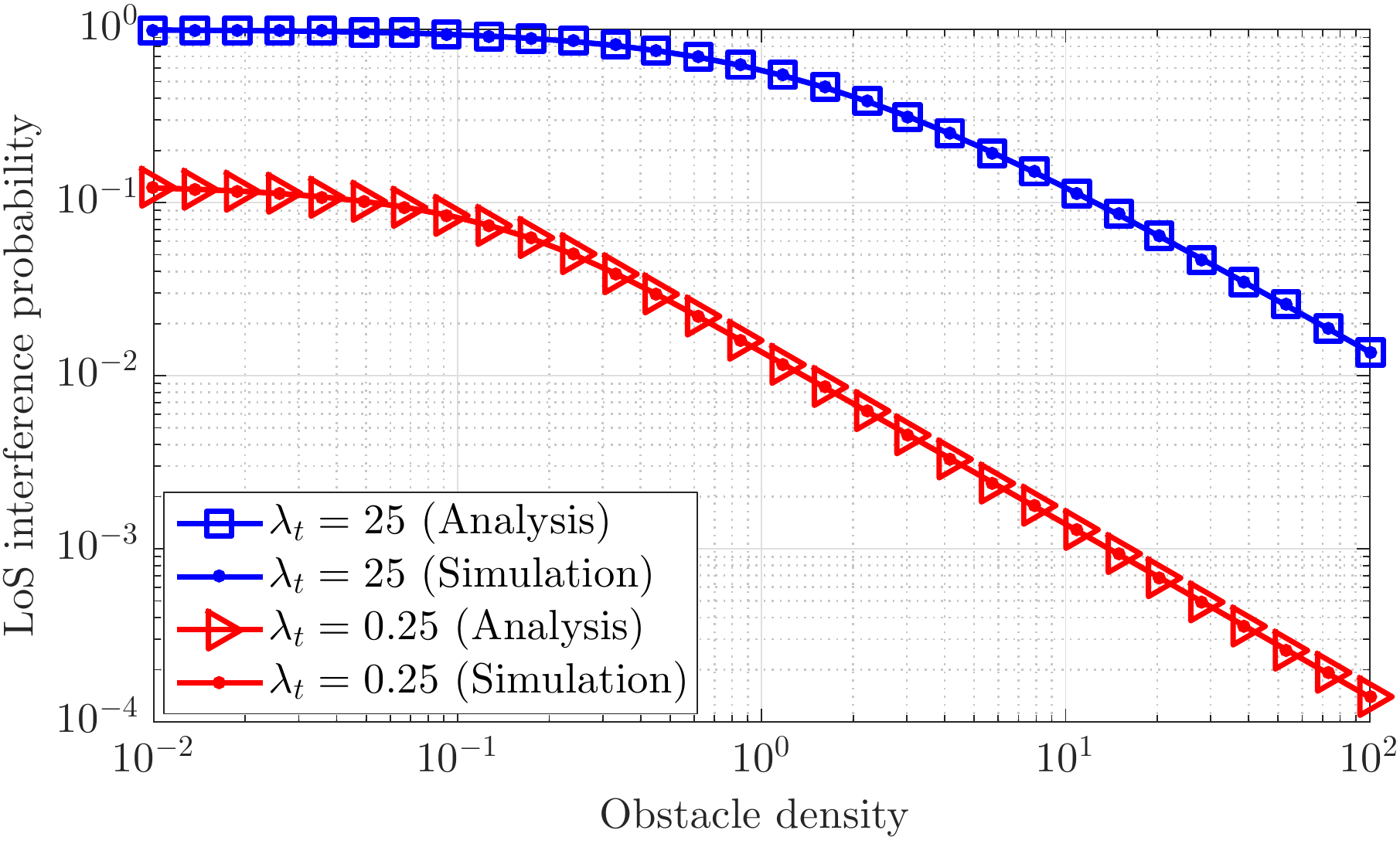}
		\label{subfig: LoSInt_SingleSector_O}
	}
	
    \caption{The probability of having LoS interference from sector $s$, ${1 \leq s \leq k-1}$, as a function of~\subref{subfig: LoSInt_SingleSector_I} link density and~\subref{subfig: LoSInt_SingleSector_O} obstacle density, as computed by Equation~\eqref{eq: LoSIwithJointPDF} and Monte Carlo simulations.}
	\label{fig: LoSInt_SingleSector}
\end{figure}
\begin{figure}[!t]
	\centering
	\includegraphics[width=\columnwidth]{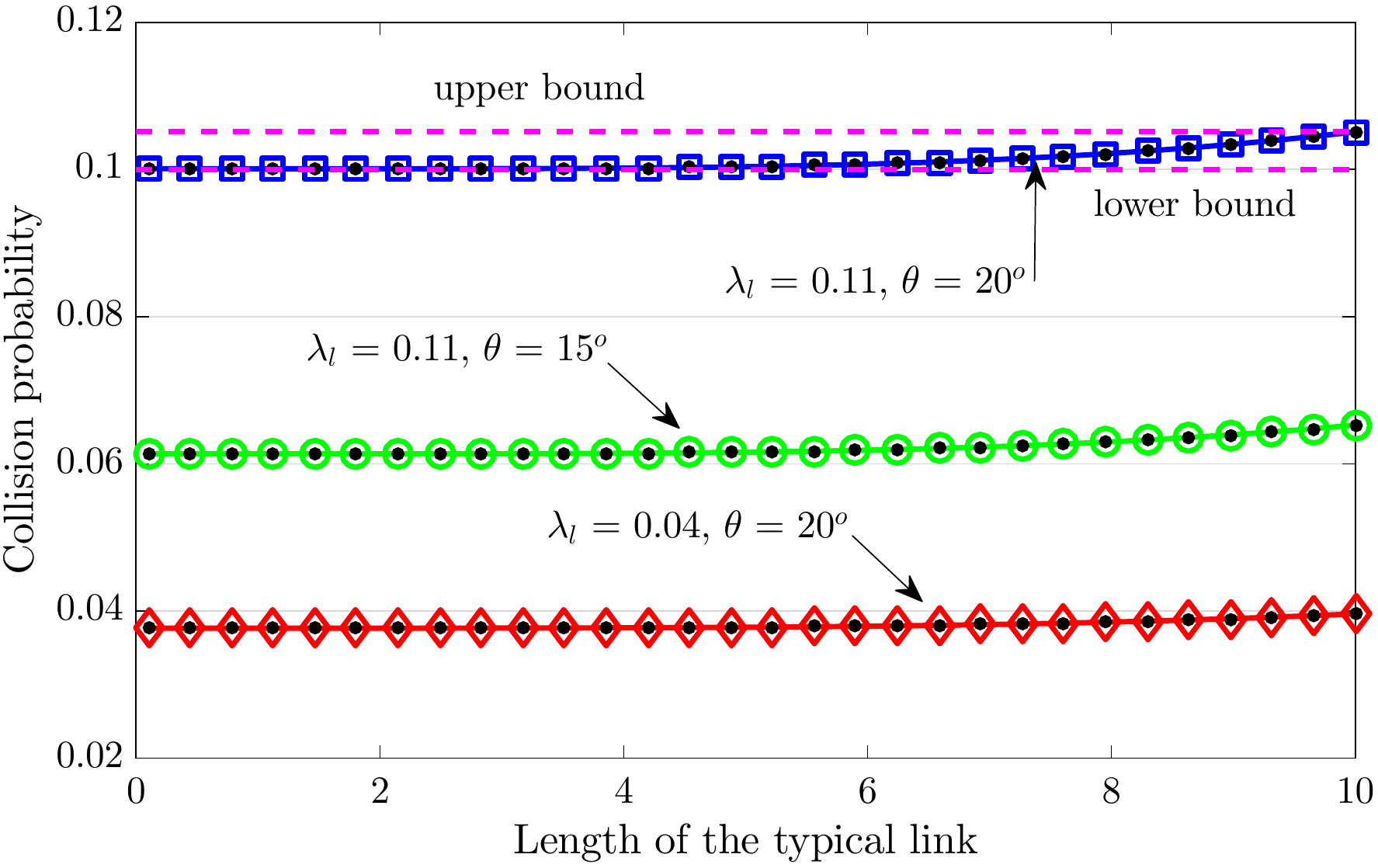}
	
    \caption{The probability of collision as a function of the length of the typical link, as computed by Equations~\eqref{eq: CollProbCond} and Monte Carlo simulations, marked by filled circles. Upper and lower bounds are computed by Equation~\eqref{eq: CollProbBounds}.}
	\label{fig: CollisionProbVsL}
\end{figure}

Fig.~\ref{subfig: LoSInt_SingleSector_I} shows the probability of having LoS interference from a given sector $s$, $1 \leq s \leq k-1$, as a function of link density $\lambda_t$. First of all, Proposition~\ref{prop: LoSI} holds for all curves. Not surprisingly, increasing the link density increases the LoS interference probability, but in a saturating manner. Also, higher obstacle density increases blockage probability, so reduces the LoS interference probability. As can be observed in the figure, for the density of 1 transmitter (interferer) in a 3x3~${\text{m}}^2$ area, increasing the density obstacles by a factor of 100, from 0.0025 to 0.25, leads to only 62\% reduction on the probability of observing an LoS interferer. To better understand the impact of obstacle density $\lambda_o$, we report the probability of having LoS interference from a given sector $s$, $1 \leq s \leq k-1$, as a function of $\lambda_o$. LoS interference probability is not too sensitive to the changes of $\lambda_o$ for small obstacle densities. However, the sensitivity increases by $\lambda_o$, leading to a very fast reduction in the LoS interference probability by a small increment of $\lambda_o$, for instance, for $\lambda_o > 1 $.


Although~\eqref{eq: LoSISec1:K-1} describes the LoS interference probability from every sector 1 to $k-1$, for sector $k$ we need to extend~\eqref{eq: LoSISec1:K-1} according to the corresponding blockage and interference models. As shown in Fig.~\ref{fig: IntRegion}, sector $k$ consists of two sub-sectors, corresponding to the distances $\left( 0, L \right]$ and $\left( L , d_{\max}\right]$. In the first sub-sector, there is no obstacle, whereas we have regular appearance of the obstacles in the second sub-sector, see Fig.~\ref{fig: IntRegion}. Following the same steps taken in Appendix~A and in Proposition~\ref{prop: LoSI}, and after some algebraic manipulations, we can derive the probability of receiving LoS interference from sector $k$ in~\eqref{eq: LoSIsubsec1}.
\begin{figure*}[!t]
\normalsize
\begin{equation}\label{eq: LoSIsubsec1}
\Pr[{\text{LoS interference from sector $k$}}] = 1 - e^{- \lambda_I A_{L}} + \frac{\lambda_I e^{\lambda_o A_{L}}}{\lambda_o + \lambda_I} \left( e^{-\left( \lambda_o + \lambda_I \right) A_{L}} - e^{-\left( \lambda_o + \lambda_I \right) A_{d_{\max}}} \right) \:.
\end{equation}
\hrulefill
\end{figure*}

\begin{prop}\label{prop: Collision Probability}
Let $\lambda_t$ and $\lambda_o$ denote the density of the interferers and obstacles per unit area. Let $\rho_a$ be the probability that an interferer is active. Consider blockage and interference models, described in Fig.~\ref{fig: IntRegion}. Let $L$, $d_{\max}$, $\theta$, and $\theta_c$ be the length of the typical link, interference range, operating beamwidth, and coherence angle, respectively. The collision probability given $L=\ell$, denoted by~$\rho_{c \mid L} (\ell)$, is given by Equation~\eqref{eq: CollProbCond} on the top of page~\pageref{eq: CollProbCond}, where $\lambda_I = \rho_a \lambda_t \theta / 2 \pi$, $A_{d_{\max}} = \theta_c d_{\max}^{2}/2$ and $A_{\ell} = \theta_c \ell^{2}/2$.
\begin{figure*}[!t]
\normalsize
\begin{equation}\label{eq: CollProbCond}
\rho_{c \mid L} \left( \ell \right) = 1 - \left( \frac{ \lambda_o + \lambda_I e^{- \left( \lambda_o + \lambda_I \right) A_{d_{\max}}}}{\lambda_o + \lambda_I} \right)^{\lceil \theta / \theta_c \rceil -1} \left( e^{- \lambda_I A_{\ell}} - \frac{\lambda_I e^{\lambda_o A_{\ell}} }{\lambda_o + \lambda_I} \left( e^{-\left( \lambda_o + \lambda_I \right) A_{\ell}} - e^{-\left( \lambda_o + \lambda_I \right) A_{d_{\max}}} \right) \right) \:.
\end{equation}
\hrulefill
\end{figure*}
\end{prop}

\begin{IEEEproof}
Given that the typical link is established, the collision probability is equal to the probability of having at least one LoS interferer, irrespective of the sectors in which the LoS interferer(s) are. To derive the collision probability, we first find its complementary, that is, the probability of having no LoS interferer in any sector. The latter is equal to complementary of the event of having collision from any sector, given by~\eqref{eq: LoSISec1:K-1} and~\eqref{eq: LoSIsubsec1}. Considering mutual independence of different sectors, the proof is straightforward.
\end{IEEEproof}

We can draw several fundamental remarks from the closed-form expression of the collision probability given by~\eqref{eq: CollProbCond}.
\begin{corollary}\label{cor: Remark1}
The collision probability, formulated in~\eqref{eq: CollProbCond}, implies the following asymptotic results:
\begin{align*}
\lambda_I \to 0 \quad\hspace{-1mm} &\Rightarrow \quad \hspace{-1mm} \rho_{c \mid L} \left( \ell \right)\to 0 \:, \\
\lambda_o \to 0 \quad\hspace{-1mm} &\Rightarrow \quad \hspace{-1mm} \rho_{c \mid L} \left( \ell \right) \to 1 - \left( e^{- \lambda_I A_{d_{\max}}}\right)^{\lceil \theta / \theta_c \rceil} \hspace{-1mm} \:, \\
\lambda_I \to \infty \:, \lambda_o < \infty \quad \hspace{-1mm}&\Rightarrow \quad \hspace{-1mm} \rho_{c \mid L} \left( \ell \right) \to 1 \:, \\
\lambda_o \to \infty \:, \lambda_I < \infty \quad \hspace{-1mm}&\Rightarrow  \quad \hspace{-1mm} \rho_{c \mid L} \left( \ell \right) \to 1 - e^{- \lambda_I A_{\ell}} \:, \\
\theta \to 0 \:, \theta = \theta_c \quad\hspace{-1mm} &\Rightarrow \quad \hspace{-1mm} \rho_{c \mid L} \left( \ell \right) \to 0 \:, \\
\theta_c \to 0 \:, \theta \gg \theta_c \quad\hspace{-1mm}  &\Rightarrow \quad \hspace{-1mm} \rho_{c \mid L} \left( \ell \right) \to 1 -  e^{- \lambda_I d_{\max}^{2} \theta/2} \:.
\end{align*}
\end{corollary}
Note that the last corollary, which can be simply proved by relaxing ceiling function in~\eqref{eq: CollProbCond} and using a Taylor expansion, is basically equivalent to assume that different interferers experience independent LoS events, as considered in~\cite{TBai2014Coverage}.
Corollary~\ref{cor: Remark1} shows asymptotic performance bounds on the conditional collision probability and provides benchmarks for the analysis.

The last step of characterizing the collision probability is taking an average of~\eqref{eq: CollProbCond} over the distribution of $L$, which is $f_L  \left( \ell \right) = 2\ell / d_{\max}^{2}$. The resulting collision probability is given by Equation~\eqref{eq: CollProbFinal} on the top of page~\pageref{eq: CollProbFinal}.
\begin{figure*}[!t]
\normalsize
\begin{equation}\label{eq: CollProbFinal}
\rho_c = 1 - \hspace{-0.5mm} \int_{\ell=0}^{d_{\max}} \! \left( \frac{ \lambda_o + \lambda_I e^{- \left( \lambda_o + \lambda_I \right) \theta_c d_{\max}^{2}/2}}{\lambda_o + \lambda_I} \right)^{\hspace{-1mm} \lceil \theta / \theta_c \rceil -1} \hspace{-2mm} \left( e^{- \lambda_I \theta_c \ell^{2}/2} - \frac{\lambda_I e^{\lambda_o A_{\ell}} }{\lambda_o + \lambda_I} \hspace{-0.8mm} \left( e^{-\left( \lambda_o + \lambda_I \right) \theta_c \ell^{2}/2} - e^{-\left( \lambda_o + \lambda_I \right) \theta_c d_{\max}^{2}/2} \right)\right) \hspace{-1mm} \frac{2\ell}{d_{\max}^{2}}  \mathrm{d}\ell .
\end{equation}
\hrulefill
\end{figure*}

\begin{prop}\label{prop: Collision Probability_Bounds}
Let $\lambda_t$ and $\lambda_o$ denote the density of the interferers and obstacles per unit area. Let $\rho_a$ be the probability that an interferer is active. Consider blockage and interference models, described in Fig.~\ref{fig: IntRegion}. Let $d_{\max}$, $\theta$, and $\theta_c$ be the interference range, operating beamwidth, and coherence angle, respectively. The collision probability is bounded as in Equation~\eqref{eq: CollProbBounds}, where $\lambda_I = \rho_a \lambda_t \theta / 2 \pi$.
\begin{figure*}[!t]
\normalsize
\begin{equation}\label{eq: CollProbBounds}
1 - \left( \frac{ \lambda_o + \lambda_I e^{- \left( \lambda_o + \lambda_I \right) \theta_c d_{\max}^{2}/2}}{\lambda_o + \lambda_I} \right)^{\lceil \theta / \theta_c \rceil } \leq  \rho_c \leq 1 - e^{- \lambda_I \theta_c d_{\max}^{2}/2} \left( \frac{ \lambda_o + \lambda_I e^{- \left( \lambda_o + \lambda_I \right) \theta_c d_{\max}^{2}/2}}{\lambda_o + \lambda_I} \right)^{\lceil \theta / \theta_c \rceil -1} \:.
\end{equation}
\hrulefill
\end{figure*}
\end{prop}

\begin{IEEEproof}
Consider~\eqref{eq: CollProbCond} and~\eqref{eq: CollProbFinal}. We first observe that the conditional collision probability given by~\eqref{eq: CollProbCond} is strictly increasing with $\ell$. Therefore, the lower and upper bounds of~\eqref{eq: CollProbFinal} are $\rho_{c \mid L} \left( 0 \right)$ and $\rho_{c \mid L} \left( d_{\max} \right)$, respectively. This completes the proof.
\end{IEEEproof}

Using simulation parameters similar to those used in Fig.~\ref{fig: LoSInt_SingleSector}, we depict $\rho_{c \mid L} \left( \ell \right)$ against $\ell$ in Fig.~\ref{fig: CollisionProbVsL}. As stated in Proposition~\ref{prop: Collision Probability_Bounds}, the conditional collision probability is an increasing function of $\ell$ with lower and upper bounds, formulated in~\eqref{eq: CollProbBounds}. First, Proposition~\ref{prop: Collision Probability} holds for all curves, and there is a perfect coincidence between numerical and analytical results. Moreover, both upper and lower bounds are tight for all examples considered in the figure, implying that the approximated closed-form expressions~\eqref{eq: CollProbBounds} can be effectively used for pessimistic/optimistic MAC layer designs, instead of the exact but less tractable expression. For the example of 1 transmitter in a 3x3~${\text{m}}^2$ area and operating beamwidth of $\theta = 20 \degree$, the maximum error due to those approximations, that is, the difference between upper and lower bounds is only 0.005. This error reduces as the operating beamwidth or the link density reduces, see Fig.~\ref{fig: CollisionProbVsL}.

In the next section, we will use the collision probability to derive several performance metrics of a mmWave ad~hoc network.

\section{Throughput and Delay Analysis}\label{sec: Performance-analysis}
The closed-form expression of the collision probability and its bounds, formulated in~\eqref{eq: CollProbCond}--\eqref{eq: CollProbBounds}, allow deriving the effective MAC layer throughput, analyzing the regime at which the network operates, highlighting inefficiency of hybrid MAC protocols of existing standards, and providing insightful discussions on the proper resource allocation and interference management protocols for future mmWave networks.

\subsection{Noise-limited or Interference-limited}
To compute per-link throughput, we note that the tagged transmitter is active with probability $\rho_a$.
Its transmission to the typical receiver at distance $L$ is successful if there is no blockage on the typical link, which occurs with probability $ e^{- \lambda_o A_{L}} $, and no collision, which occurs with probability $\left( 1 - \rho_{c \mid L} \left( \ell \right) \right)$. Therefore, the conditional probability of successful transmission in a slot given $L=\ell$ is
\begin{equation}\label{eq: MACthroughputCond}
\rho_{s \mid L} \left( \ell \right) = \rho_a e^{- \lambda_o A_{\ell}} \left( 1 - \rho_{c \mid L} \left( \ell \right) \right) \:.
\end{equation}
Let $r_{_{\text{S-ALOHA}}}$ be the average MAC throughput of slotted ALOHA. Assuming transmission of one packet per slot, the average per-link throughput is equal to the average successful transmission probability, hence
\begin{align}\label{eq: MACthroughputFinal0}
r_{_{\text{S-ALOHA}}} &= \int_{\ell=0}^{d_{\max}} \! \rho_{s \mid L} \left( \ell \right) f_L\left( \ell \right) \, \mathrm{d}\ell \nonumber \\
&= \int_{\ell=0}^{d_{\max}} \! \rho_a e^{- \lambda_o A_{\ell}} \left( 1 - \rho_{c \mid L} \left( \ell \right) \right) \frac{2\ell}{d_{\max}^{2}}\, \mathrm{d}\ell \:,
\end{align}
where $f_L(\ell)$ is the distribution function of the link length. Since $\rho_{s \mid L} \left( \ell \right)$ is strictly decreasing with $\ell$, upper and lower bounds of $r_{_{\text{S-ALOHA}}}$, are $\rho_{s \mid L} \left( 0 \right)$ and $\rho_{s \mid L} \left( d_{\max} \right)$, given by Equation~\eqref{eq: MACThroughputBounds}.
\begin{figure*}[!t]
\normalsize
\begin{equation}\label{eq: MACThroughputBounds}
\rho_a e^{- \left( \lambda_o + \lambda_I \right) \theta_c d_{\max}^{2}/2} \left( \frac{ \lambda_o + \lambda_I e^{- \left( \lambda_o + \lambda_I \right) \theta_c d_{\max}^{2}/2}}{\lambda_o + \lambda_I} \right)^{\lceil \theta / \theta_c \rceil -1} \leq  r_{_{\text{S-ALOHA}}} \leq \rho_a \left( \frac{ \lambda_o + \lambda_I e^{- \left( \lambda_o + \lambda_I \right) \theta_c d_{\max}^{2}/2}}{\lambda_o + \lambda_I} \right)^{\lceil \theta / \theta_c \rceil } \:.
\end{equation}
\hrulefill
\vspace*{4pt}
\end{figure*}

For a given $\rho_a$, the throughput is uniquely determined by the collision probability. It follows that we can study the collision probability, instead of the throughput, to identify the operating regime. By definition, we are in the \emph{noise-limited} regime if the collision probability is too small for given density of the obstacles, density of the transmitters, and operating beamwidth, among the main parameters. On the other hand, if there is at least one LoS interferer, which limits the throughput performance of the typical link, we are in the \emph{interference-limited} regime. This suggests the following conclusion. A mmWave network with directional communication exhibits a \emph{transitional behavior}, that is, a transition from a noise-limited regime to an interference-limited regime. This transition depends on the density of interferers and obstacles, transmission probability, operating beamwidth, transmission powers, coherence angle, and also the MAC protocol.

\begin{figure}[!t]
	\centering
    \subfigure[]{
	\includegraphics[width=\columnwidth]{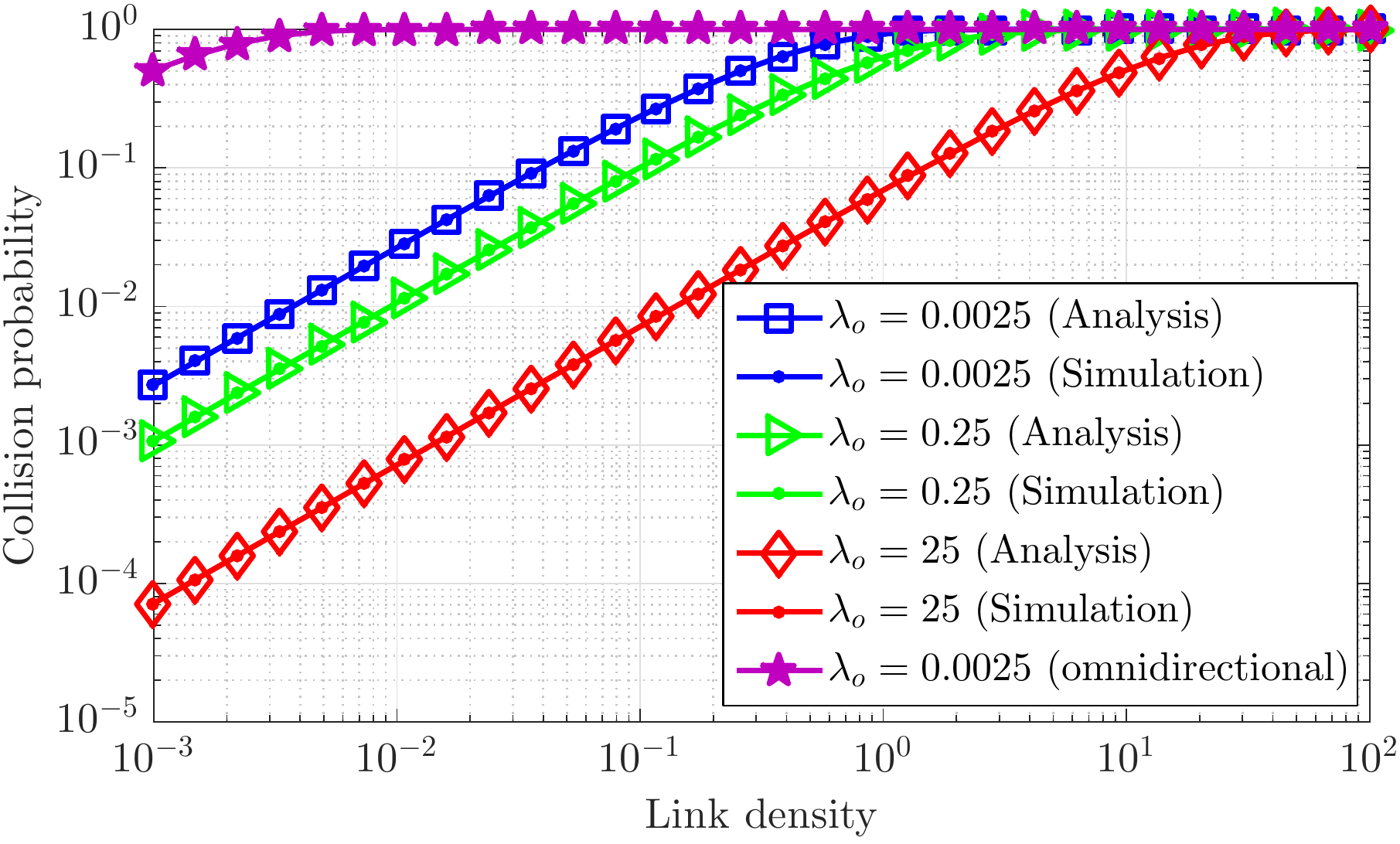}
		\label{subfig: CollisionGivenL_I}
	}
	\subfigure[]{
	\includegraphics[width=\columnwidth]{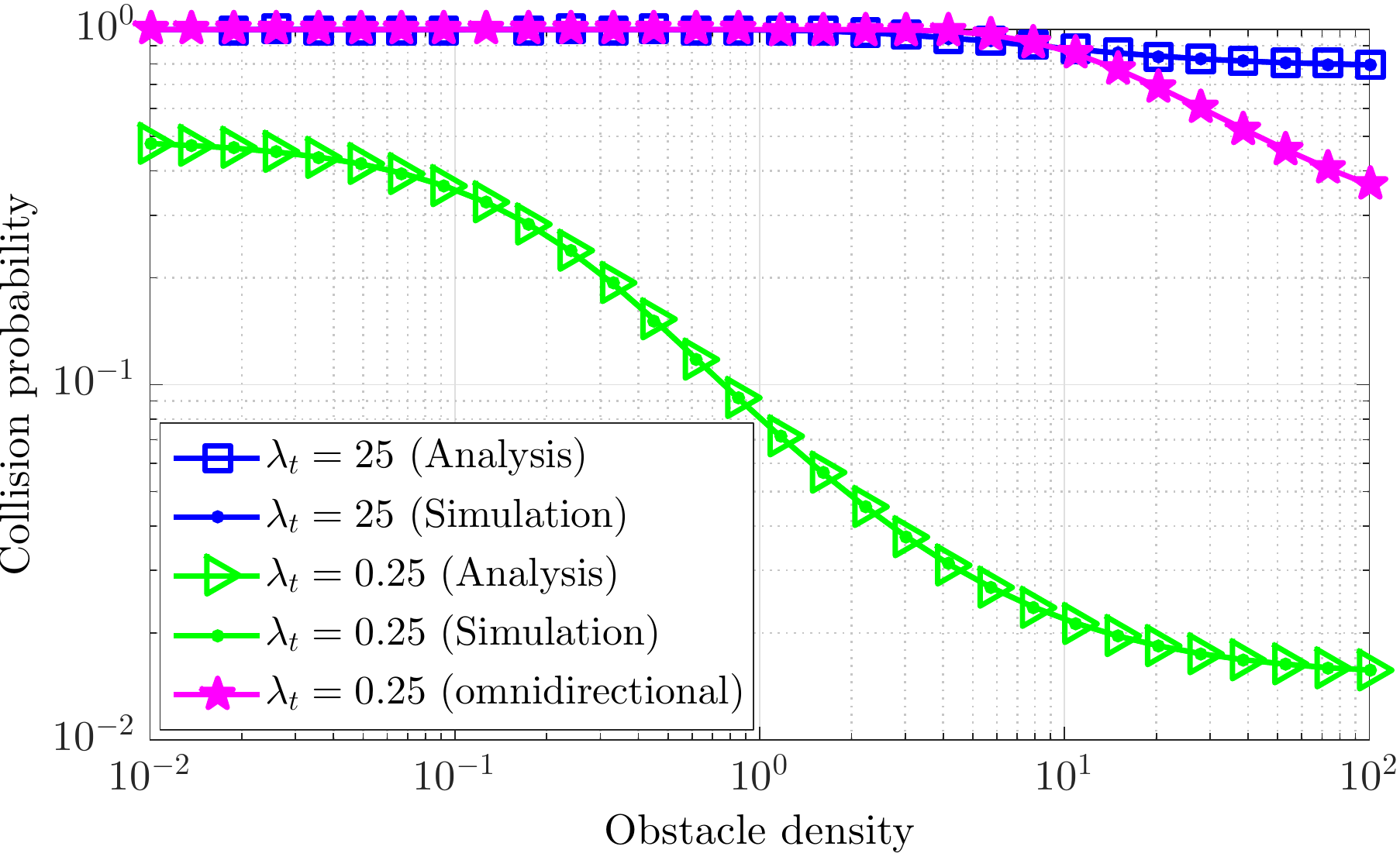}
		\label{subfig: CollisionGivenL_O}
	}
	
    \caption{The probability of collision as a function of~\subref{subfig: CollisionGivenL_I} link density and~\subref{subfig: CollisionGivenL_O} obstacle density. The length of the typical link is 5~m.}
	\label{fig: CollisionProb}
\end{figure}
We use the same simulation parameters as of Fig.~\ref{fig: LoSInt_SingleSector} to investigate the collision probability as a function of $\lambda_t$ and $\lambda_o$, depicted in Fig.~\ref{fig: CollisionProb}. From Fig.~\ref{subfig: CollisionGivenL_I}, collision probability is not negligible even for a modest size mmWave network. For instance, for 1 transmitter in a 3x3~${\text{m}}^2$ area and 1 obstacle in a 20x20~${\text{m}}^2$ area, the collision probability is as much as 0.26. Increasing the density of the obstacles to 1 obstacle in a 3x3~${\text{m}}^2$ area, which is not shown in Fig.~\ref{subfig: CollisionGivenL_I} for the sake of clarity, the collision probability reduces to 0.17, which is still high enough to invalidate the assumption of being in a noise-limited regime. This conclusion becomes even more clear in Fig.~\ref{subfig: CollisionGivenL_O}, where the green curve represents a collision probability as high as 0.48 for not so dense WPANs (1 transmitter in a 2x2~${\text{m}}^2$).
Moreover, as can be observed in all curves of Fig.~\ref{subfig: CollisionGivenL_I}, there is a transition from the noise-limited regime to the interference-limited one. For benchmarking purposes, we also simulate a network with omnidirectional communications. Fixing all other parameters, we only increase the transmission power to achieve the same interference range as the corresponding directional communications and investigate the collision probability. As shown in Fig.~\ref{fig: CollisionProb}, traditional networks with omnidirectional communications always experience an interference-limited regime without any transitional behavior.

In this paper, we have considered only the LoS interference. Upon existence of some reflectors with sufficiently large reflection coefficients such as tinted glass~\cite{Zhao201328Ghz}, besides LoS aligned unintended transmitters, some other unintended transmitters in deafness/blokage condition may now cause collision at the typical receiver. This is equivalent to increase the density of the potential interferers from $\rho_a \lambda_t \theta/ 2 \pi$  to $\rho_a \lambda_t \theta/ 2 \pi + \lambda_n $, where $\lambda_n$ corresponds to the non-LoS interferers and is a function of the reflector process (density, average size, and reflection coefficient), transmitter and obstacle densities, and operating beamwidth.\footnote{We may need independence between the density of the LoS interferers and that of the non-LoS interferers for the analysis. Such independence does not hold in general, since a LoS interferer may also have a first order reflected path to the typical receiver. However, due to directivity and blockage of mmWave networks, neglecting such independence introduces negligible error into the interference model, as we have extensively investigated in~\cite{Shokri2015WhatIs}.} Given $\lambda_n >0$, the higher density of the potential interferers shifts all curves of Fig.~\ref{subfig: CollisionGivenL_I} to the left, indicating that the typical receiver experiences the same collision probability for smaller values of the transmitter density $\lambda_t$. Mathematical modeling of $\lambda_n$ is the subject of our future studies.

\subsection{Proper Resource Allocation Protocol}
In this subsection, we compare the MAC layer throughput of a single link, area spectral efficiency (network throughput normalized to the network size), and delay performance of slotted ALOHA to those of TDMA in a mmWave network. We define delay as the difference between the time a new packet is inserted to the transmission queue of the transmitter and the time it is correctly received at the receiver. We also numerically investigate the performance of CSMA and CSMA/CA to make rigorous conclusions about the resource allocation protocols suitable for mmWave networks.

\begin{figure}[!t]
	\centering
	\includegraphics[width=0.99\columnwidth]{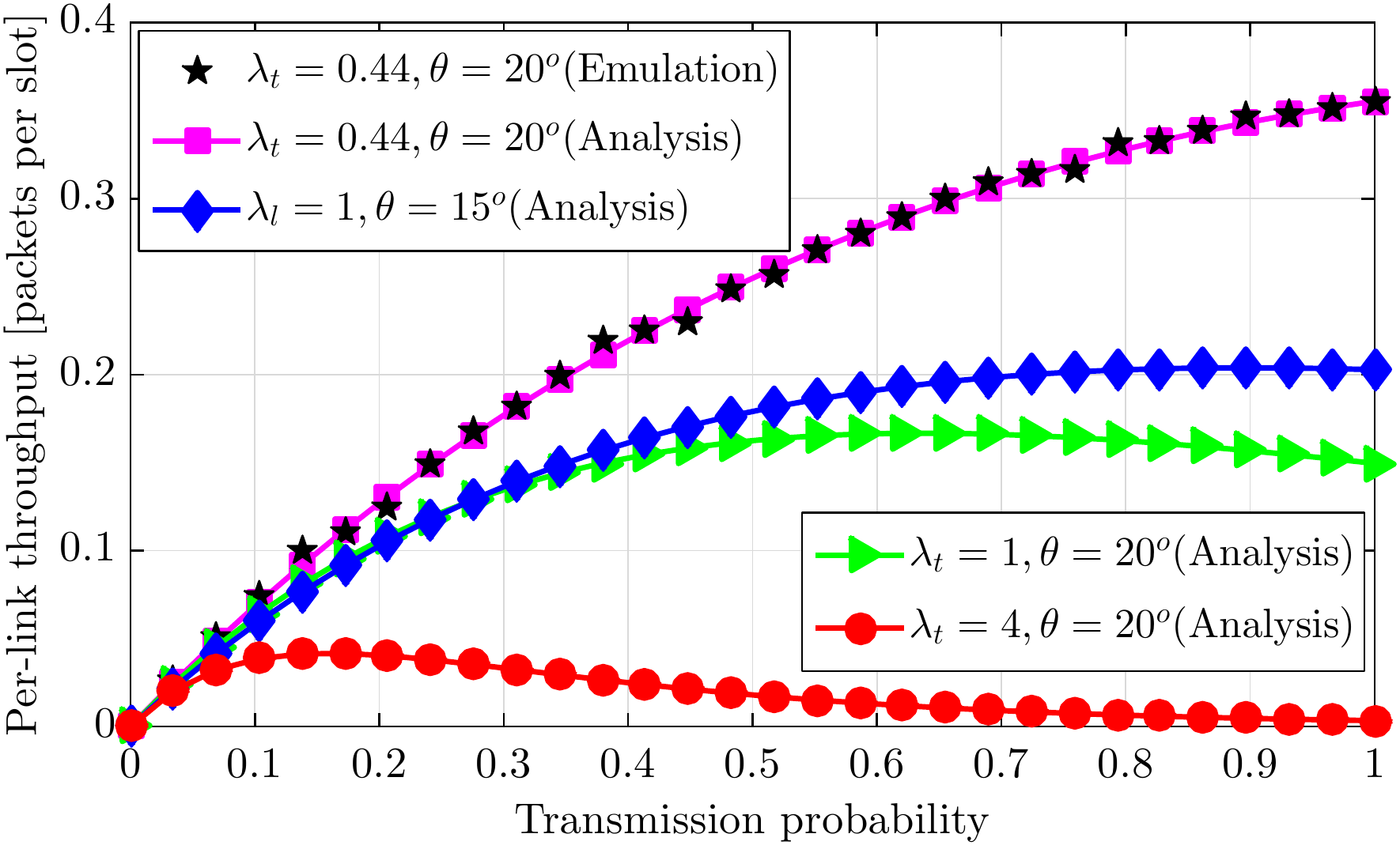}
	
    \caption{Per-link throughput against transmission probability $\rho_a$, as computed by the emulator and by Equation~\eqref{eq: MACthroughputFinal0}. The obstacle density is $\lambda_o = 0.11$ per unit area. The coherence angle in analytical figures is $\theta_c = 5 \degree$.}
	\label{fig: EffectiveMACThroughput}
\end{figure}

Per-link throughput of slotted ALOHA is derived in~\eqref{eq: MACthroughputFinal0}. To evaluate its area spectral efficiency (ASE), we consider a large region with area $A$. The number of transmitters (links) inside this region is $1 + n_t$, where $n_t$ follows a Poisson distribution with mean $A \lambda_t$. We assume that, at each transmission attempt, and regardless of the number of retransmissions suffered, each packet collides with constant and independent probability $\rho_c$ (given by Equation~\eqref{eq: CollProbFinal}), which is also independent of the number of transmitters. This is a common assumption in the throughput analysis of IEEE~802.15.4~\cite{park2009generalized,pollin2008performance} and IEEE~802.11~\cite{bianchi2000performance,hui2005unified,malone2007modeling,garetto2008modeling}, which can be extended to the general case using similar approach taken in~\cite{jang2012ieee}. Also, we show the validity of this assumption in Figs.~\ref{fig: EffectiveMACThroughput} and~\ref{subfig: ASE-LinkDensity}. With this independence assumption, the network throughput is $\left( 1 + n_t\right) r_{_{\text{S-ALOHA}}} $, leading to an average network throughput of $\left( 1 + A \lambda_t\right) r_{_{\text{S-ALOHA}}}$. Thus, ASE of slotted ALOHA, denoted by ${\mathrm{ASE}}_{\text{S-ALOHA}}$, is
\begin{align}\label{eq: AreaSpecEffic}
{\mathrm{ASE}}_{\text{S-ALOHA}} &= \frac{1 + A \lambda_t}{A} r_{_{\text{S-ALOHA}}} \nonumber \\
&= \frac{1 + A \lambda_t}{A} \hspace{-1mm} \int_{\ell=0}^{d_{\max}} \! \frac{2 \ell \rho_a}{d_{\max}^{2}}  e^{- \lambda_o A_{\ell}} \left( 1 - \rho_{c \mid L} \left(\ell \right) \right)  \mathrm{d}\ell \:,
\end{align}
which can be tightly approximated by $\lambda_t r_{_{\text{S-ALOHA}}}$ if $A\lambda_t~\gg~1$. This condition holds for networks with high density of the transmitters (high $\lambda_t$) or for those with large size (high $A$).

We can also use the derived collision probability to analyze the delay performance of slotted ALOHA. In the following, we only show the main steps and leave the exact calculations for future studies. Let $\rho_s$ denote the probability of successful transmission, derived in~\eqref{eq: MACthroughputCond} and~\eqref{eq: MACthroughputFinal0}. Let $n_r$ be the number of retransmissions in the typical link until successful reception. $n_r$ can be accurately approximated by a geometric distribution~\cite{Yang2003Delay}, that is,
\begin{equation*}
\Pr [n_r=n_{r_0}] =  \rho_s \left( 1 - \rho_s  \right)^{n_{r_0}}.
\end{equation*}
Let $w_i$ be the contribution of $i$-the transmission/retransmission on the total delay, where $w_0$ is the delay due to initial transmission. Each $w_i$ contains round-trip propagation, packet transmission, and backoff delays~\cite{Yang2003Delay}. Then, the delay is $\sum_{i=0}^{n_r}{w_i}$. Detailed analysis of the delay is out of the scope of this work, and we use Monte Carlo simulations to find the delay performance.

Unlike slotted ALOHA, TDMA protocol activates only one link at a time, regardless of the number of links. This guarantees a collision-free communication. We derive the throughput of a link and ASE of TDMA in the following proposition:
\begin{prop}\label{prop: TDMAthroughputPerformance}
Consider the blockage model, described in Fig.~\ref{fig: IntRegion}. Let $\lambda_o$ be the density of the obstacles, $\theta_c$ be the coherence angle, and $d_{\max}$ be the interference range. Consider a typical link. Let $A$ denote the area over which TDMA regulates the transmissions of $1 + n_t$ links, including the typical link, where $n_t$ is a Poisson random variable with density $\lambda_t$ per unit area. Average per-link throughput under TDMA scheduler is
\begin{equation}\label{eq: TDMALinkThroughput}
r_{_{\text{TDMA}}} =  \left( \frac{1 - e^{-\lambda_t A}}{\lambda_t A} \right) \left( \frac{1-e^{-\lambda_o A_{d_{\max}}}}{\lambda_o A_{d_{\max}}} \right) \:.
\end{equation}
where $A_{d_{\max}} = \theta_c d_{\max}^2/2$. Moreover, ASE under TDMA scheduler is
\begin{equation}\label{eq: TDMA-ASE}
{\mathrm{ASE}}_{\text{TDMA}} =  \frac{1-e^{-\lambda_o A_{d_{\max}}}}{A \lambda_o A_{d_{\max}}} \:.
\end{equation}
\end{prop}

\begin{IEEEproof}
A proof is given in Appendix~B.
\end{IEEEproof}

\begin{corollary}\label{cor: 1}
Consider~\eqref{eq: MACthroughputFinal0} and~\eqref{eq: TDMALinkThroughput}. We have
\begin{equation*}
\lim\limits_{\lambda_t \to 0}{r_{_{\text{S-ALOHA}}}} = \lim\limits_{\lambda_t \to 0}{r_{_{\text{TDMA}}}} = \frac{1-e^{-\lambda_o A_{d_{\max}}}}{ \lambda_o A_{d_{\max}}} \:.
\end{equation*}
\end{corollary}
Corollary~\ref{cor: 1} implies that, even without any interferer in the network ($\lambda_t \to 0$), per-link throughput of 1 packet per slot is not achievable if $\lambda_o > 0$. The main reason is the non-zero probability of having obstacle(s) on the typical link.

\begin{corollary}\label{cor: 2}
Upper bounds on the throughput performance of TDMA scheduler are
\begin{equation*}
r_{_{\text{TDMA}}} \leq \frac{1 - e^{-A \lambda_t}}{A \lambda_t} \:, \qquad
{\mathrm{ASE}}_{\text{TDMA}} \leq \frac{1}{A} \:,
\end{equation*}
which can be achieved if $\lambda_o A_{d_{\max}} \to 0$.
\end{corollary}

\begin{IEEEproof}
We first note that $\left( 1 - e^{-x} \right) /x$ is strictly decreasing for any $x>0$, and that $x=\lambda_o A_{d_{\max}} >0$. Therefore, \eqref{eq: TDMALinkThroughput} and \eqref{eq: TDMA-ASE} can be upper bounded by letting $x \to 0^{+}$. Using $\mathop{\lim\hspace{3mm}}\limits_{x \to 0^{+}}{\left( 1 - e^{-x} \right) /x} \to 1$, we conclude the proof.
\end{IEEEproof}

\begin{corollary}\label{cor: 3}
Consider Corollary~\ref{cor: 2}. Per-link throughput under TDMA scheduler goes to zero as the average number of links in the network $A \lambda_t$ grows large. Moreover, ASE of TDMA protocol goes to zero as the network size $A$ grows large.
\end{corollary}

Corollaries~\ref{cor: 2} and~\ref{cor: 3} explicitly show the inefficiency of TDMA protocol to share resources among massive number of devices in a mmWave network. Besides poor throughput performance, the delay of TDMA increases with the number of activate transmitters, as a transmitter should wait more to access the channel~\cite{benvenuto2011principles}. In the following, we numerically compare the throughput and delay performance of slotted ALOHA to those of TDMA.

To validate the blockage model as well as the assumption of independence of $\rho_c$ and the number of transmitters, introduced in the throughput analysis, we build an ns3-based mmWave emulator. We consider a random number of aligned mmWave links (aligned transmitter-receiver pairs) on 2D space, all operating with the same beamwidth at 60~GHz. The transmitters and receivers are uniformly distributed in a 10x10~${\text{m}}^2$ area. We also generate a random number of obstacles with density $\lambda_o$ and uniformly distribute them in the environment. The obstacles are in the shape of lines with random orientations and their lengths are uniformly distributed between 0 and 1~m. Every transmitter generates traffic with constant bit rate (CBR) 384~Mbps, the size of all packets is 10~kB, time slot duration is 50~$\mu$s, transmission rate is 1 packet per slot (link capacity around 1.5~Gbps), the transmitters have infinite buffer to save and transmit the packets, and the emulation time is 1~s.
We also simulate CSMA/CA of IEEE~802.11ad~\cite{802_11ad}, where each transmitter sends a request-to-send (RTS) before channel access and the corresponding receiver sends back clear-to-send (CTS) to reserve the channel. A sequence of random backoffs may be executed by every transmitter to solve possible collisions. To increase the robustness, IEEE~802.11ad adopts peak transmission rate of 27.7~Mbps for signaling messages. Moreover, every device should wait for an SIFS duration ($2.5$~$\mu$s) before sending every RTS, CTS, and ACK, and should wait for a DIFS duration ($5.5$~$\mu$s) before every regular data frame. We consider 30~Bytes for RTS, CTS, and ACK messages.

We first start with a mmWave network operating with slotted ALOHA protocol.
Fig.~\ref{fig: EffectiveMACThroughput} shows the per-link throughput as a function of transmission probability. First of all, there is an excellent match between the results obtained from the emulator and those from Equation~\eqref{eq: MACthroughputFinal0} with $\theta_c = 5 \degree$, which confirms the validity of both blockage model and the independence assumption. Moreover, for relatively not dense networks, for instance, 1 transmitter in a 1.5x1.5~${\text{m}}^2$ area ($\lambda_t = 0.44$), increasing the transmission probability is always beneficial, as the multiuser interference level is small enough that activating more links will not substantially reduce the average throughput of a link but increases the number of time slots over which the link is active. As the link density increases, higher collision probability introduces a tradeoff on increasing the transmission probability and reducing the interference. In a very dense network, for instance, with $\lambda_t = 4$, we should adopt a very small transmission probability to maximize the per-link throughput.

\begin{figure}[!t]
	\centering
\subfigure[]{
	\centering
		\includegraphics[width=\columnwidth]{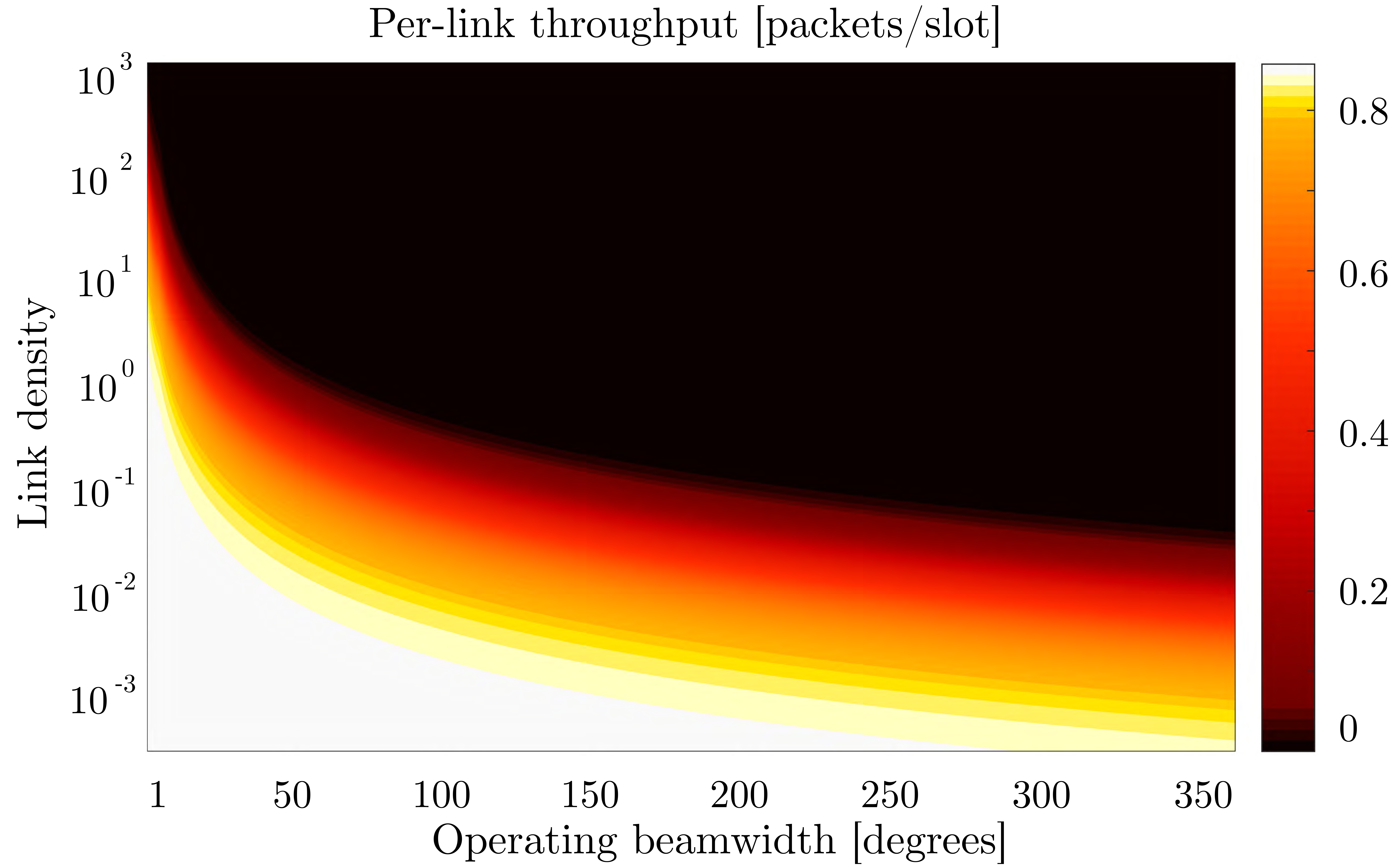}
	   \label{subfig: Heat_LinkThr_beamwidth}
}
\subfigure[]{
	\centering
		\includegraphics[width=\columnwidth]{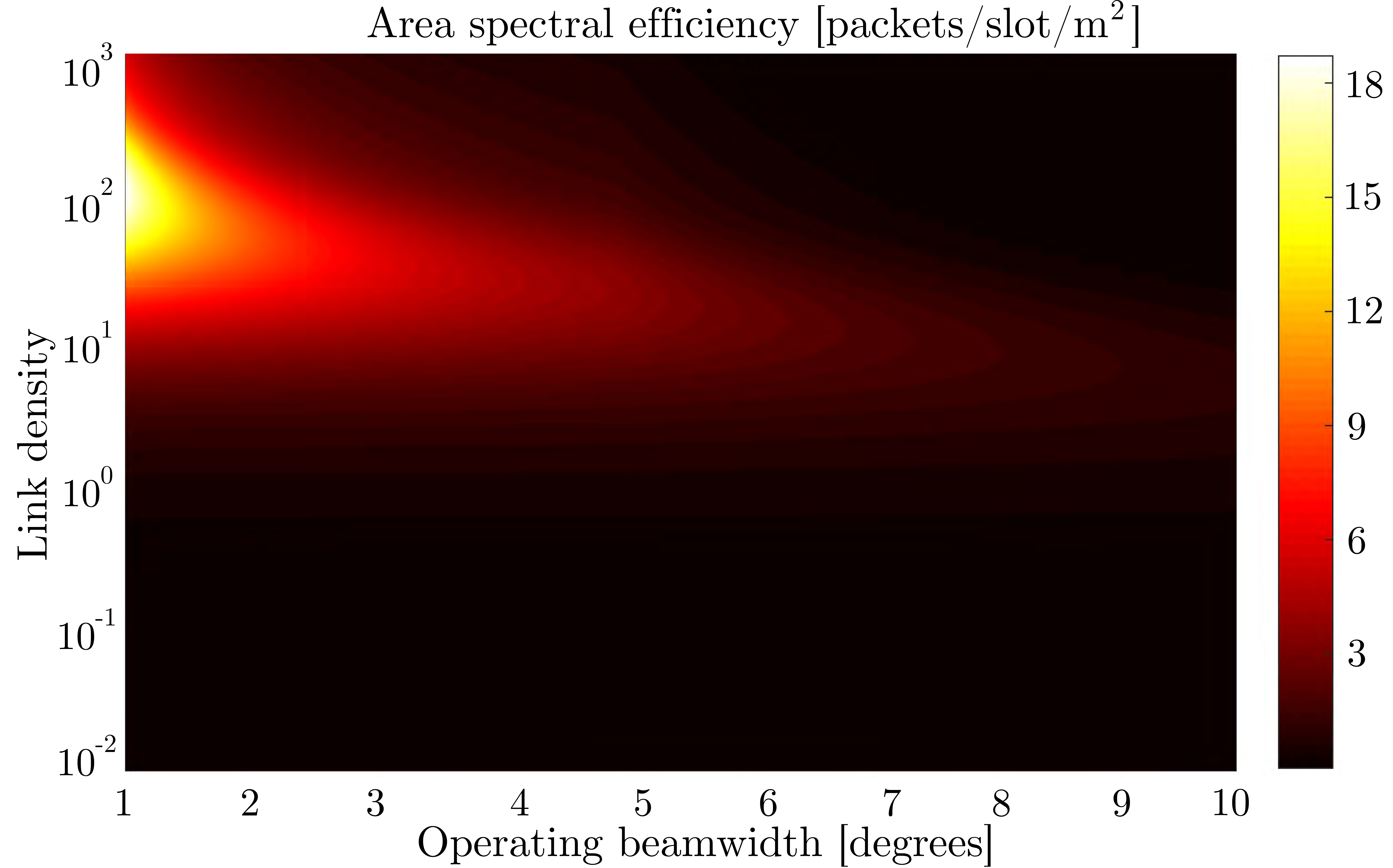}
	   \label{subfig: Heat_ASE_beamwidth}
}

\caption{Achievable regions of \subref{subfig: Heat_LinkThr_beamwidth} per-link throughput and \subref{subfig: Heat_ASE_beamwidth} area spectral efficiency of slotted~ALOHA with $\rho_a = 1$.}
\label{fig: HeatDiagram}
\end{figure}
Fig.~\ref{fig: HeatDiagram} illustrates the achievable regions of per-link throughput and ASE of slotted ALOHA with $\rho_a = 1$ and $\lambda_o = 0.11$. Brighter colors correspond to higher values. For instance, with operating beamwidth $\theta = 50 \degree$ and on average 2 transmitters in a square meter, a per-link throughput of 0.5 packets per slot is not achievable. To achieve that, we should reduce either the operating beamwidth or the link density (or equivalently the transmission probability). The per-link throughput is always less than 1 packet per slot due to blockage on the typical link, see Corollary~\ref{cor: 1}. From Fig.~\ref{fig: HeatDiagram}, there is a tradeoff between operating beamwidth and link density. To maintain a certain level of per-link throughput or ASE, we can either increase the operating beamwidth or the link density. Furthermore, these figures confirm that without alignment overhead, mmWave networks benefit from narrower operating beamwidth and denser deployment. However, as mentioned in~\cite{Shokri2015Beam}, adopting extremely narrow beams is not throughput optimal in general due to the alignment overhead.

\begin{figure}[!t]
	\centering
	\subfigure[]{
	\includegraphics[width=0.985\columnwidth]{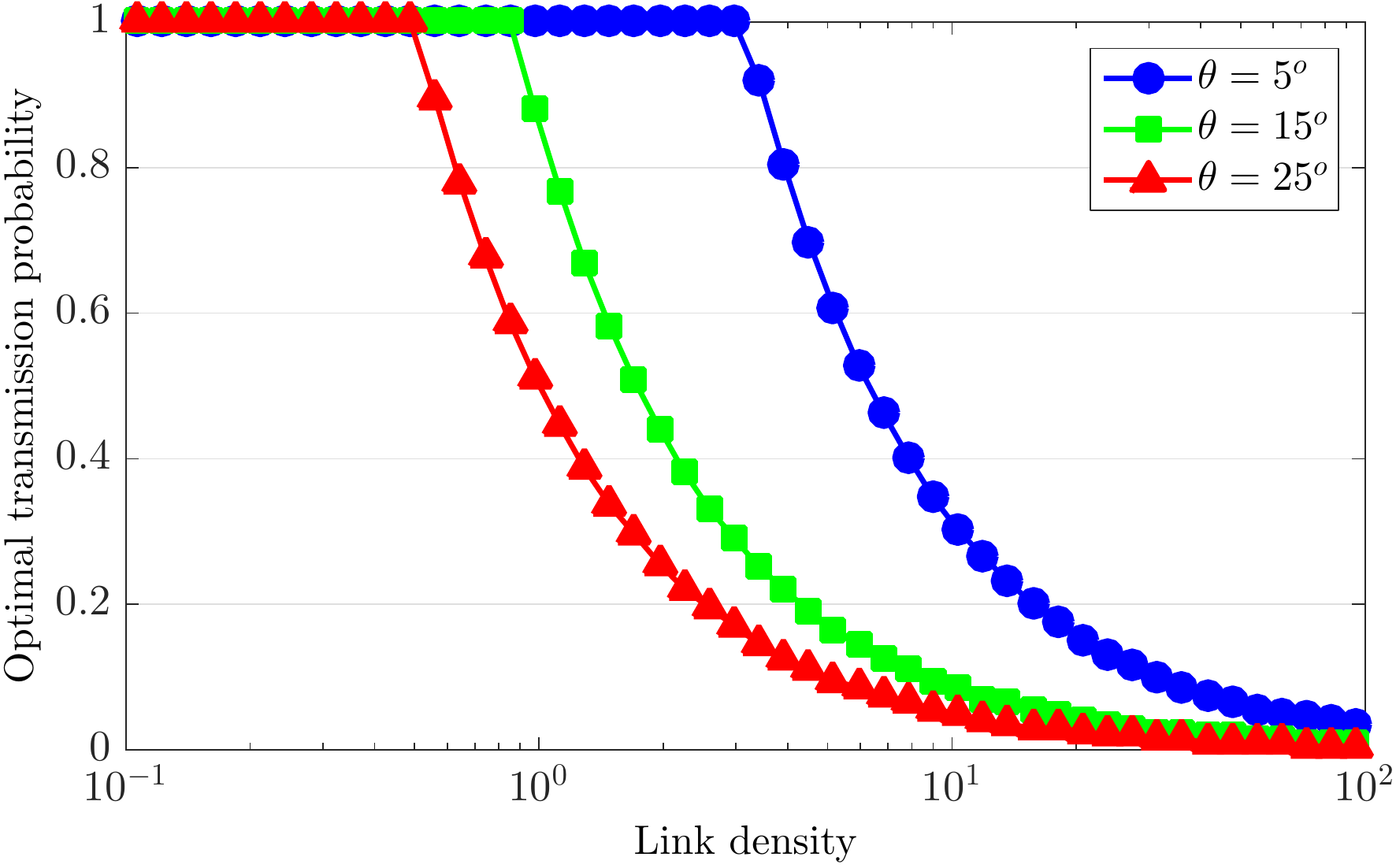}
		\label{subfig: OptimMACThr_OptimlTxProb}
	}
    \subfigure[]{
	\includegraphics[width=0.985\columnwidth]{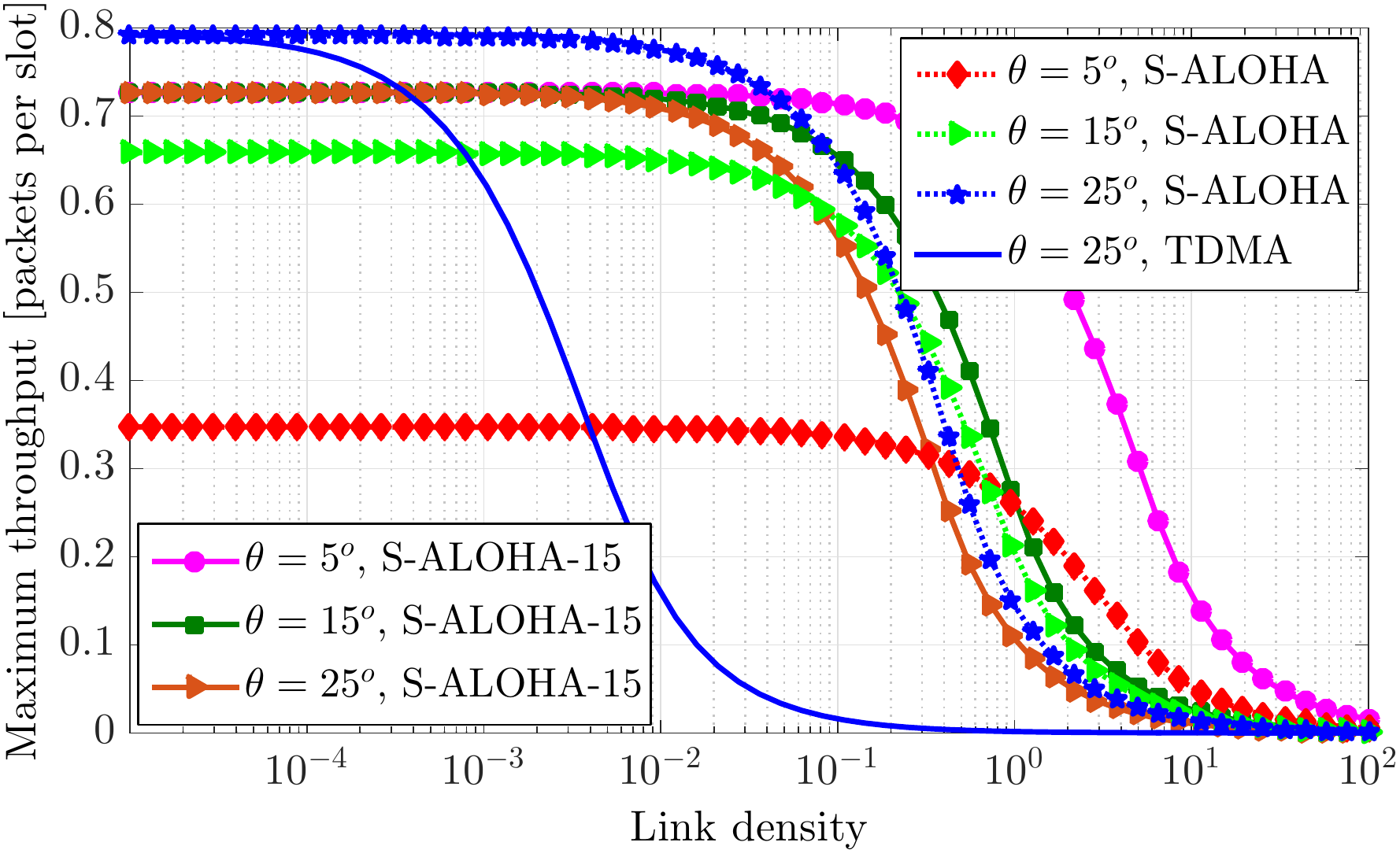}
		\label{subfig: OptimMACThr_MaxThr}
	}
	
    \caption{~\subref{subfig: OptimMACThr_OptimlTxProb} The optimal transmission probability and~\subref{subfig: OptimMACThr_MaxThr} the maximum per-link throughput against link density. ``S-ALOHA'' stands for slotted ALOHA, and ``S-ALOHA-15'' refers to slotted ALOHA with $d_{\max} = 15$.}
	\label{fig: OptimMACThr}
\end{figure}

\begin{figure}[!t]
	\centering
	\subfigure[]{
		\includegraphics[width=\columnwidth]{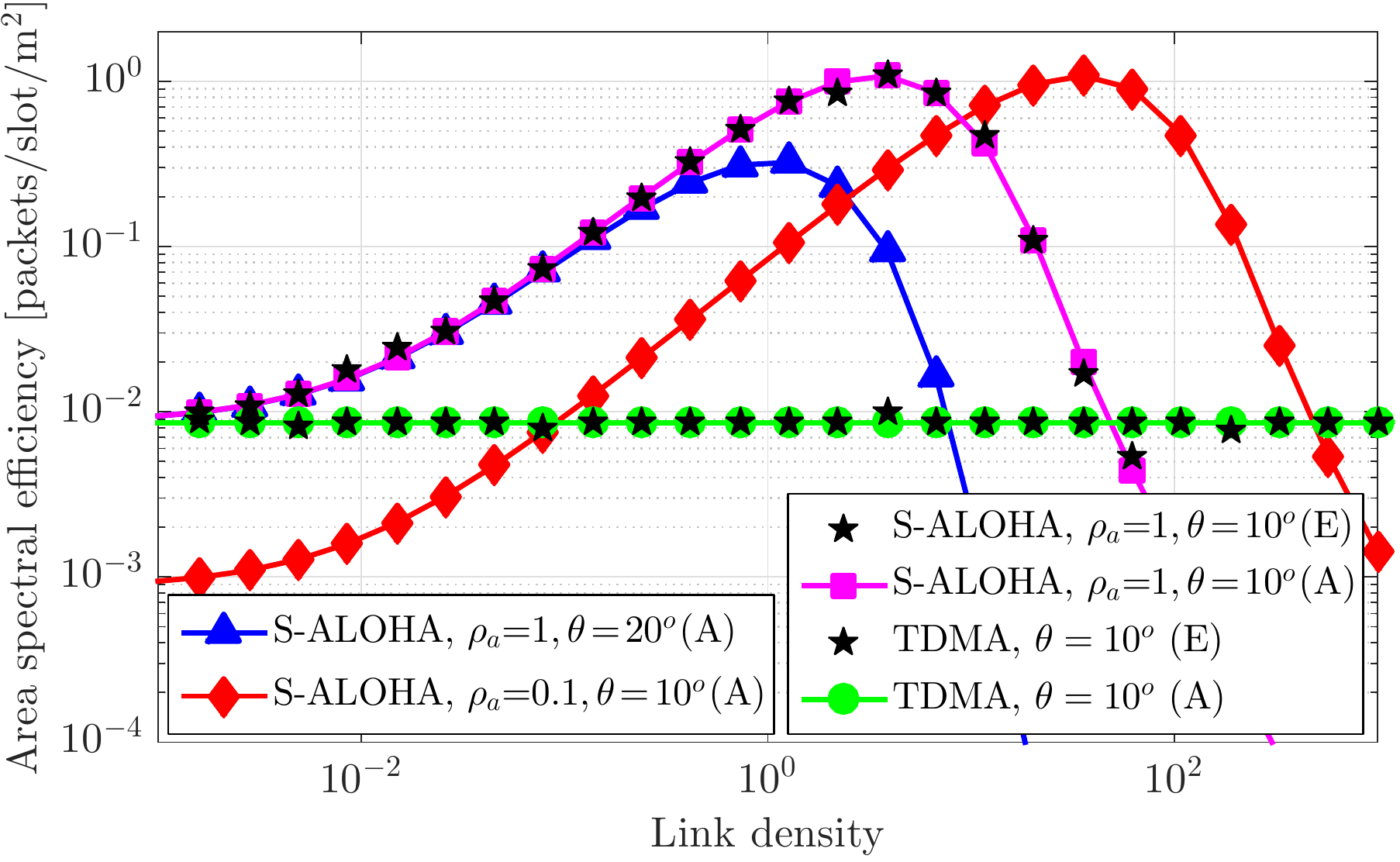}
		\label{subfig: ASE-LinkDensity}
	}
	\subfigure[]{
		\includegraphics[width=\columnwidth]{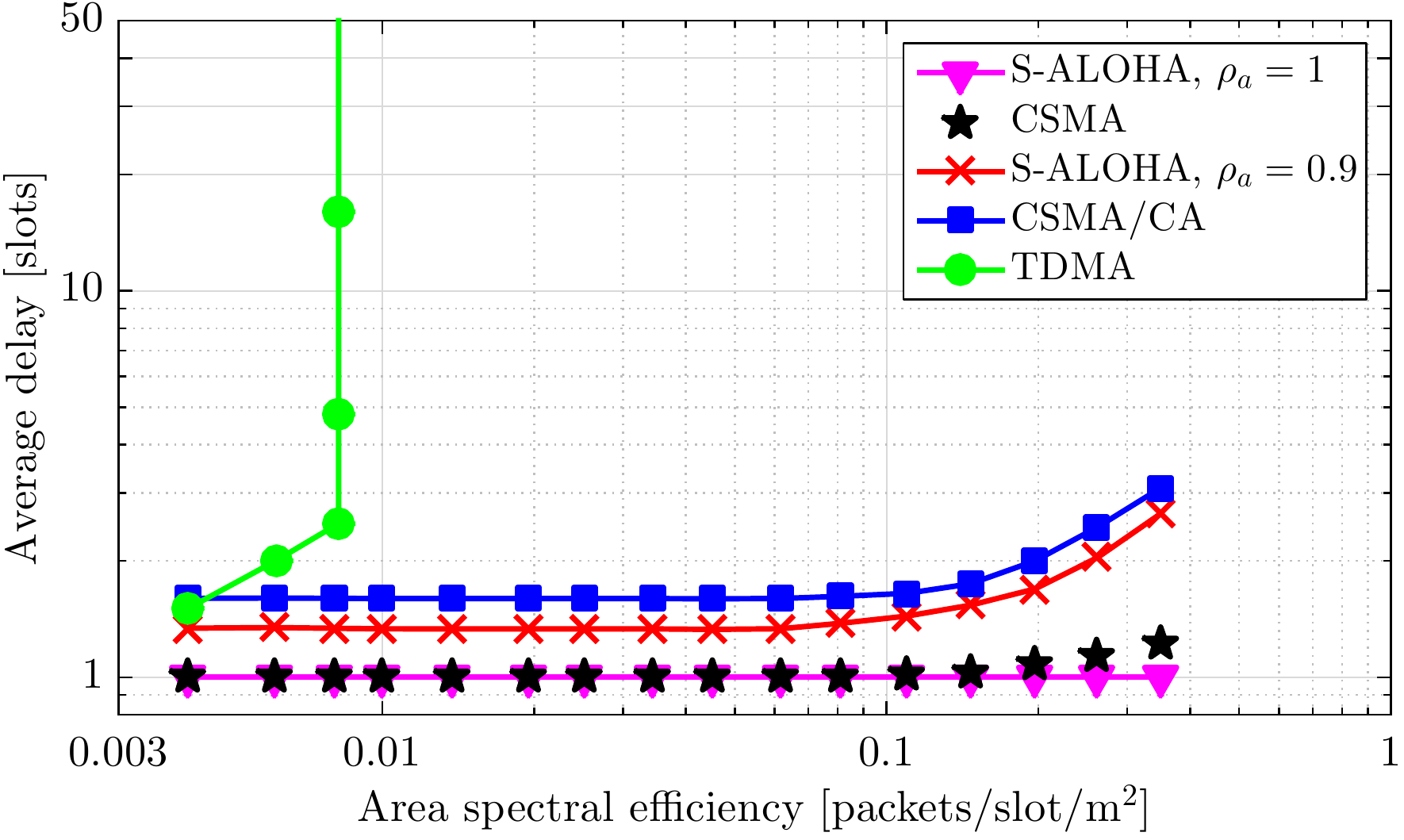}
		\label{subfig: ASE-Delay}
	}
	
	\caption{Area spectral efficiency and delay performance of slotted ALOHA and those of TDMA. Area size is 10x10~${\text{m}}^2$. Operating beamwidth in \subref{subfig: ASE-Delay} is 10$^{\degree}$. Different points of \subref{subfig: ASE-Delay} represent different link densities (up to 2 links per square meter). The obstacle density is $\lambda_o = 0.25$ per unit area. Slotted ALOHA provides substantially higher ASE with lower delay. These performance gains may improve with the number of links.}
	\label{fig: ASE-Delay}
\end{figure}

Fig.~\ref{fig: OptimMACThr} shows the behavior of the optimal transmission probability of slotted ALOHA (that maximizes per-link throughput) as a function of link density $\lambda_t$ and operating beamwidth $\theta$. Thanks to this figure, we can explicitly answer why there is a throughput degradation, as observed in~\cite{Shokri2015Beam}, if we activate all links at the same time and under which conditions such a degradation will disappear. From Fig.~\ref{subfig: OptimMACThr_OptimlTxProb}, in many cases, the optimal transmission probability is 1, implying that we can simply activate all links and still achieve the maximum MAC throughput. In fact, negligible multiuser interference of those cases makes the performance of one of the simplest collision-based resource allocation scheme (slotted ALOHA) almost equivalent to the optimal collision-free resource allocation scheme (STDMA) with much lower signaling and computational overheads. However, as the operating beamwidth or the link density increases, we should think of more intelligent resource allocation strategies as the mmWave network may transit to the interference-limited regime. This further invalidates the generality of the noise-limited mmWave networks and indicates that we may adopt a very small transmission probability to decrease the contention level in an ultra dense mmWave network.
Fig.~\ref{subfig: OptimMACThr_MaxThr} demonstrates the maximum throughput of a link in slotted ALOHA, associated with the optimal transmission probability. In the first set of curves of this figure, we fixed the interference range $d_{\max}$ to 15, whereas in the second set we let $d_{\max}$ change according to $\theta$, see~\eqref{eq: d_max}. Fixing either link length or $d_{\max}$ (only the latter is depicted for the sake of clarity in the figure), the per-link throughput in slotted ALOHA will monotonically increase with decreased $\theta$. That is because, according to~\eqref{eq: CollProbCond} and~\eqref{eq: MACthroughputCond}, narrower beams reduce the collision probability, so increase $\rho_{s \mid L} \left( \ell\right)$, leading to a higher average $r_{_{\text{S-ALOHA}}}$. Therefore, with fixed $d_{\max}$, we always have \emph{lower beamwidth higher throughput} rule. However, if we do not manually fix $d_{\max}$ (e.g., by changing the transmission power), lower $\theta$ causes another effect, namely extended length at which a link can be established. This extended communication range, in turn, increases the blockage probability and may consequently reduce the average throughput. In other words, two parameters with a non-trivial interplay affect the average throughput: blockage and collision. For sparse networks, the reduced blockage probability due to a higher $\theta$ dominates the increased collision probability, and we can observe \emph{higher beamwidth higher throughput} rule. However, higher link density introduces more collisions to the network and highlights the impact of the collision term on the average throughput. After a critical link density, the reduced blockage probability due to a higher $\theta$ cannot compensate for the increased collision probability, so we can observe \emph{lower beamwidth higher throughput} rule.

As illustrated in Fig.~\ref{subfig: OptimMACThr_OptimlTxProb}, slotted ALOHA significantly outperforms TDMA. The main reason is that TDMA realizes an orthogonal use of time resources, irrespective of the collision level, whereas slotted ALOHA re-uses all the time resources and benefits from the spatial gain. This gain leads to 390\% and 4270\% throughput enhancements over TDMA for the cases of 1 transmitter in a 10x10~${\text{m}}^2$ and in a 3x3~${\text{m}}^2$ area with $\theta=25 \degree$, respectively. Note that, from Fig.~\ref{subfig: OptimMACThr_OptimlTxProb}, the optimal transmission probability is 1 in both cases, further highlighting simplicity of the corresponding slotted ALOHA.
Per-link throughput in TDMA is strictly decreasing with density of the transmitters, whereas that of slotted ALOHA remains almost unchanged as long as the collision term, shown in~\eqref{eq: MACthroughputCond} and \eqref{eq: MACthroughputFinal0}, is almost negligible.
As stated in Corollary~\ref{cor: 3}, the throughput of TDMA goes to zero very fast. Although slotted ALOHA shows the same asymptotic zero throughput behavior, it has much slower rates of convergence to this asymptotic point. Considering any arbitrary small $\zeta$ for the per-link throughput, from Fig.~\ref{subfig: OptimMACThr_MaxThr}, the per-link throughput of both TDMA and slotted ALOHA become lower than $\zeta$ for sufficiently large $\lambda_{t}$; however, slotted ALOHA reaches that point with almost two orders of magnitude more links in the network (e.g., see $\zeta = 0.1$), indicating its efficiency on handling massive wireless access in mmWave networks.

We use the developed mmWave emulator to find ASE and the average delay performance. Fig.~\ref{subfig: ASE-LinkDensity} illustrates ASE of slotted ALOHA and that of TDMA as a function of link density. Again, there is a perfect coincidence between the analytical results obtained from Equations~\eqref{eq: AreaSpecEffic} and~\eqref{eq: TDMA-ASE} and those of the emulator.
Increasing the number of links in the network does not affect ASE of TDMA.\footnote{Note that TDMA can increase the network throughput if individual transmitters do not have enough payload to occupy the whole time slot. In this case, TDMA divides one long time slot to smaller pieces, each for one transmitter, leading to higher channel utilizations. However, in this paper, we have assumed that every packet of a transmitter requires one time slot, so the TDMA channel is already saturated if the transmitters have always packets to transmit.} The average network throughput of TDMA is slightly lower than one packet per slot, and it achieves the upper bound if the obstacle density goes to zero, see Corollary~\ref{cor: 2}. Slotted ALOHA with transmission probability $\rho_a = 1$ provides the highest ASE, which is firstly increasing with the link density and then shows a strictly decreasing behavior once the throughput loss, due to the collision term, overweighs the throughput enhancement due to the first term of~\eqref{eq: AreaSpecEffic}. For the example of $\rho_a = 1$ and $\theta = 10 \degree$, the optimal density of transmitters that maximizes ASE is, on average, 3.5 transmitters per square meter. This example number indeed means that, from the perspective of ASE, mmWave networks benefit from dense deployment. Slotted ALOHA with $\rho_a = 0.1$ outperforms that with $\rho_a=1$ in ultra dense WPANs ($\lambda_t>9$ in Fig.~\ref{subfig: ASE-LinkDensity}), as lower transmission probability leads to fewer active links. Moreover, narrower beams provide higher ASE.

Fig.~\ref{subfig: ASE-Delay} reports ASE and the corresponding delay of TDMA, slotted ALOHA , CSMA, and CSMA/CA. Slotted ALOHA with transmission probability 1 is the best strategy from both ASE and delay perspectives. It introduces only one slot delay, that is, a packet transmission time. However, if a link observes a collision at its first transmission attempt, it cannot successfully transmit anymore, as we do not have any randomness in the transmission time (e.g., with random backoff techniques). To solve this issue, we can use slotted ALOHA with transmission probability less than 1 (e.g., 0.9), but at the expense of extra delay with exponential growth at very high network throughput (equivalently high ASE). Note that this delay is still around 2 slots for a very dense mmWave WPAN with 2 transmitters in a unit area, in the example considered. Moreover, slotted ALOHA with transmission probability 0.9 avoids transmissions of each link with probability 0.1, even for a sparse mmWave network with negligible multiuser interference, introducing unnecessary extra delay compared to slotted ALOHA with transmission probability 1. CSMA/CA can address the problems of slotted ALOHA, though introduces a serious problem in mmWave networks: massive overhead of proactive collision avoidance procedure. Virtual channel reservation with the traditional RTS/CTS mechanism imposes a substantial delay in the channel access and therefore significantly reduces the network throughput (and thus ASE). The main reason is the significant mismatch between transmission rates of the data (up to 6.7~Gbps in IEEE~802.11ad) and control packets (up to 27.7~Mbps in IEEE~802.11ad). For instance, sending one 10~kB data packet with CSMA/CA under the assumption of no collision at the receiver requires around 28~$\mu$s channel reservation (1 RTS, 1 CTS, 2 SIFS, and 1 DIFS) plus 50~$\mu$s data transmission (assuming data rate of 1.5~Gbps). This leads to around 64\% channel utilization, which will be further reduced to 28\% for 6.7~Gbps data rate. This initial channel reservation delay is visible in Fig.~\ref{subfig: ASE-Delay} at very low ASE values, where instead of having 1 slot delay to send a data packet, the total delay is around 1.6 slots. Altogether, with almost negligible hidden and exposed node problems in mmWave networks~\cite{Shokri2015Licentiate} and comparatively very low transmission rate of control messages, the use of the conventional collision avoidance procedure becomes less justifiable. For ultra dense mmWave networks, not shown in Fig.~\ref{subfig: ASE-Delay}, the hidden and exposed node problems may start again to be non-negligible and reduce the network throughput, justifying the use of CSMA/CA. CSMA with random backoff, as an alternative approach, not only can solve the problems of slotted ALOHA without introducing any extra delay to the interference-free links, also can efficiently handle a few collisions that may happen in mmWave networks without using costly collision avoidance procedure. Detailed comparison of CSMA and CSMA/CA is out of the scope of this paper and left for future studies.
Finally, with TDMA, the delay increases with the link density with no significant network throughput gain. Considering traffic generation rate of this example, which is 0.25 of the link capacity, the network will be saturated roughly with 4 links in the environment, and further increasing the number of links will not improve the network throughput, but reduces the time share of every link and consequently reduces the average throughput of a link. Note that with a fixed packet generation rate, \emph{effective link capacity} (links capacity multiplied by its time share) in TDMA reduces with the number of links in the network, so the queues of the transmitter may become unstable. The delay in slotted ALOHA is not significantly affected by the total number of transmitters; rather it depends on the number of transmitters in the collision domain of the typical receiver --those that can cause collision to the typical receiver. This number may be much smaller than the total number of transmitters in mmWave networks, thanks to directionality and blockage. Furthermore, due to the time-reuse, the effective link capacity of slotted ALOHA is significantly higher than that of TDMA.
Superior throughput and delay performance of slotted ALOHA is due to the spatial gain. As the network goes to the noise-limited regime, spatial gain and consequently throughput/delay gains improve.

\subsection{Collision-aware Hybrid MAC}
Although slotted ALOHA may outperform TDMA in terms of throughput/delay, the latter guarantees collision-free communication, which is necessary for specific applications. The transitional behavior of interference in mmWave networks indicates inefficacy of the existing standards and suggests dynamic incorporations of both contention-based and contention-free phases in the resource allocation. The current mmWave standards such as IEEE~802.15.3c and IEEE~802.11ad adopt similar resource allocation approaches as those developed for the conventional interference-limited networks, e.g., by IEEE~802.15.4~\cite{park2009generalized}. In particular, they introduce a contention-based phase mainly to register channel access requests of the devices inside the mmWave network. These requests are served on the following contention-free phase, called service period in IEEE~802.11ad~\cite{802_11ad}. In fact, though some data packets with low QoS requirements may be transmitted in the contention-based phase, the network traffic is mostly served in the contention-free phase irrespective of the network operating regime. Instead, we can (and should) leverage the transitional behavior of mmWave networks to dynamically serve the network traffic partially on the contention-based and partially on the contention-free phase, according to the actual network operating regime. More specifically, a data transfer interval,\footnote{Data transfer interval is introduced in IEEE~802.11ad~\cite{802_11ad}. Similar interval in the superframe of IEEE~802.15.3c consists of the contention access period and the channel time allocation period~\cite{802_15_3c}.} that is, a set of consecutive time slot over which devices will be scheduled for data transmission, can consist of a two phases:
\begin{itemize}
  \item \emph{phase 1:} a distributed contention-based resource allocation, which is more suitable for the noise-limited regime.
  \item \emph{phase 2:} a centralized contention-free resource allocation, which is more suitable for the interference-limited regime.
\end{itemize}
While all devices can contend to access the channel in the first phase, only devices with collided packets or those with a common receiver will be scheduled on the second phase. For a noise-limited regime, automatically, most of the traffics will be served on the first phase due to negligible multiuser interference. In an interference-limited regime, however, many links may register their collisions --so their channel access requests-- to be scheduled on the following contention-free phase. Using flexible phase duration, adjusted according to the collision level of the networks, we can realize an on-demand use of the inefficient contention-free phase, improve the network throughput (especially as the network goes to the noise-limited regime), and also guarantee collision-free communications.

Directional communications in mmWave networks substantially alleviates the hidden and exposed node problems~\cite{Shokri2015Licentiate}, diminishing the advantages of the collision avoidance procedure of CSMA/CA. The transitional behavior of interference, along with high probability of having no multiuser interference at many receivers, further challenges proactive execution of the collision avoidance procedure as it is already adopted by current mmWave standards. The transitional behavior of interference in mmWave networks raises a fundamental question if a mmWave transmitter still needs to regularly send expensive and inefficient control signals to avoid possible collisions, irrespective of the actual network operating regime. This suggests the investigation of new contention-based protocols with an on-demand collision avoidance capability.



\section{Concluding Remarks}\label{sec: Conclusion}
Millimeter wave (mmWave) communication systems use directional transmission and reception to compensate for severe channel attenuation and for high noise power. This narrow-beam operation significantly reduces multiuser interference footprint, promising a significant spatial gain that is largely ignored in the resource allocation approach of current mmWave standards. In this paper, we derived a tractable closed-form expression for collision probability in a mmWave ad~hoc network operating under slotted ALOHA. This derivation allowed investigation of the MAC layer throughput of a mmWave network, as a function of the transmitter density, obstacle density, transmission probability, operating beamwidth, and transmission power, among the main parameters. Comprehensive analysis revealed that mmWave networks exhibit a transitional behavior from a noise-limited network to an interference-limited network. This transitional behavior of interference necessitates novel frameworks of collision-aware hybrid MAC, containing both contention-based and contention-free phases with adaptive phase duration.
Mathematical and numerical analysis of the per-link throughput, area spectral efficiency (network sum throughput divided by the network size), and the delay performance, indicated inefficacy of TDMA in mmWave network with small multiuser interference. Instead, slotted ALOHA efficiently leverages spatial gain and provides substantially higher throughput with lower average delay. These gains increase with the number of links in the network, making the contention-based strategies more justifiable in massive mmWave access scenarios. Moreover, the results highlighted a significant performance drop due to the conventional proactive execution of collision avoidance procedure, which imposes unnecessary overhead to many links that experience no multiuser interference. Inspired by these results, the transitional behavior of interference in mmWave networks may necessitate new collision-aware hybrid CSMA/CA-TDMA MAC for future mmWave standards, where not only the TDMA phase should be realized in an on-demand fashion, but also the collision avoidance procedure of CSMA/CA should be reactively executed to maximize the throughput and delay performance of mmWave networks. The on-demand transmission of the collision avoidance messages can be further extended to the on-demand transmissions of many other control messages to minimize the signaling overhead. This imposes a thorough modification of the traditional MAC design principles in future mmWave networks.

This paper introduced the notion of coherence angle, proposed a novel blockage model for mmWave networks, provided a new framework to analyze the performance of mmWave networks with blockage and deafness, derived closed-form expressions for the collision probability in mmWave networks along with per-link throughput and area spectral efficiency of slotted ALOHA as well as those of TDMA, clarified the collision level in a mmWave network with uncoordinated transmitters, discovered the transitional behavior of interference in mmWave networks, identified the inefficiency of the resource allocation approaches of the existing mmWave standards, and raised the necessity of on-demand contention-free resource allocation.

In this study, we did not consider the alignment (beam-searching) overhead~\cite{Shokri2015Beam}. That is, the time required for finding the best set of beams at the transmitter and at the receiver that maximizes the link budget. Boosting link budget and suppressing interference in mmWave systems with narrow-beam operation come at the expense of more complicated connection management (establishment, maintenance, and recovery) strategies. Upon missing the established channel, either due to appearance of a random obstacle or loss of precise beamforming information (e.g., due to mobility/channel change), the transmitter and receiver should trigger a time consuming alignment procedure to find another channel. Adopting narrower beams increases execution frequency of the alignment procedure. Therefore, the alignment overhead may be overwhelming and dictate the overall performance of the network, especially for networks with high mobility~\cite{shokri2015mmWavecellular}. Introducing the alignment overhead in the performance evaluation is an interesting future direction.

\section*{Appendix~A: \break Proof of Lemma~\ref{prop: jointPDF}}\label{appen: searching-overhead}
In this appendix, we find the probability of having at least one LoS interferer given the number of interferers $n_I \geq 1$ and the number of obstacles $n_o \geq 1$. We have the following lemma:

\begin{lemma}\label{lemma: Lemma2}
Let $\{X_{1},X_{2},\ldots,X_{n_I}\}$ be a set of $n_I$ i.i.d. continuous random variables with CDF $F_X(x) = x^2/d_{\max}^2$ and PDF $f_X(x)=2x/d_{\max}^2$, where $n_I$ is a zero-truncated Poisson random variable with density $\lambda_I$. Define ${X_{(1)} = \min \{X_{1},X_{2},\ldots,X_{n_I}\}}$. Given $n_I = n \geq 1$, the joint PDF of $X_{(1)}$ and $n_I$ is given by Equation~\eqref{eq: jointPDF-Lemma1} on the upper part of page~\pageref{eq: jointPDF-Lemma1}.
\begin{figure*}[!t]
\normalsize
\begin{equation}\label{eq: jointPDF-Lemma1}
f_{X_{(1)},n_I}\left( X_{(1)} = x , n_I = n | n \geq 1 \right) = \frac{2nx}{d_{\max}^2}\left( 1 - \frac{x^2}{d_{\max}^2} \right)^{n-1} \frac{e^{-\lambda_I}}{1-e^{-\lambda_I}}\frac{\lambda_{I}^{n}}{n!} \:.
\end{equation}
\hrulefill
\end{figure*}
\end{lemma}

\begin{IEEEproof}
We define $k$-order statistic of $\{X_{i}\}_{1}^{n_I}$, denoted by $X_{(k)}$, as $k$-th smallest value of $\{X_{i}\}_{1}^{n_I}$~\cite{david1970order}. Therefore, ${X_{(1)} = \min \{X_{1},X_{2},\ldots,X_{n_I}\}}$ is the first order statistic whose PDF is~\cite{david1970order}
\begin{equation}\label{eq: CDF-Lemma1}
f_{X_{(1)}}\left( x \right) = n f_X \left( x \right) \Big( 1 - F_X \left( x \right) \Big)^{n-1} \:.
\end{equation}
Noting that $n_I = n \geq 1$ is a random variable with zero-truncated Poisson distribution, thus~\cite{johnson2005univariate}
\begin{equation}\label{eq: zero-truncated}
\Pr\left[ n_I = n | n \geq 1\right]  = \frac{e^{-\lambda_I}}{1-e^{-\lambda_I}}\frac{\lambda_{I}^{n}}{n!} \:.
\end{equation}
Now, replacing PDF and CDF of random variables $\{X_{i}\}_{1}^{n_I}$ in~\eqref{eq: CDF-Lemma1} and multiplying the result by~\eqref{eq: zero-truncated}, we have~\eqref{eq: AppB1}. This concludes the proof.
\begin{figure*}[!t]
\normalsize
\begin{align}\label{eq: AppB1}
f_{X_{(1)},n_I}\left( X_{(1)} = x , n_I = n | n \geq 1 \right)  & = f_{X_{(1)}| n_I}\left( x | n_I = n , n \geq 1 \right) \Pr\left[ n_I = n | n \geq 1 \right] = \frac{2nx}{d_{\max}^{2}}\left( 1 - \frac{x^2}{d_{\max}^{2}} \right)^{n-1} \frac{e^{-\lambda_I}}{1-e^{-\lambda_I}}\frac{\lambda_{I}^{n}}{n!} \:.
\end{align}
\hrulefill
\end{figure*}
\end{IEEEproof}

Due to mutual independence of the interferer and obstacle processes, and using Lemma~\ref{lemma: Lemma2}, we obtain~\eqref{eq: AppB2}.
\begin{figure*}[!t]
\normalsize
\begin{equation}\label{eq: AppB2}
f_{X_{(1)},Y_{(1)},n_I,n_o} \left( x , y, n, m | n,m \geq 1 \right)  = f_{X_{(1)},n_I} \left( X_{(1)} = x , n_I = n | n \geq 1 \right) f_{Y_{(1)},n_o} \left( Y_{(1)} = y, n_o = m | m \geq 1 \right) \:.
\end{equation}
\hrulefill
\end{figure*}
Applying Lemma~\ref{lemma: Lemma2} to $f_{X_{(1)},n_I} \left( X_{(1)} = x , n_I = n | n \geq 1 \right)$ and $f_{Y_{(1)},n_o} \left( Y_{(1)} = y, n_o = m | m \geq 1 \right)$, the first part of Lemma~\ref{prop: jointPDF} is straightforward. All we need to do is substituting the average number of interferers and obstacles in a sector $\lambda_I A_{d_{\max}}$ and $\lambda_o A_{d_{\max}}$ into~\eqref{eq: jointPDF-Lemma1}.

The next step is finding the probability of having at least one LoS interferer given $n_I \geq 1  , n_o \geq 1$, which we denote by $I_{\mathrm{LoS}}$. We have~\eqref{eq: AppB3}, where $(\star)$ follows from the Taylor series of the exponential function. This completes the proof of Lemma~\ref{prop: jointPDF}. \begin{figure*}[!t]
\normalsize
\begin{align}\label{eq: AppB3}
I_{\mathrm{LoS}} & = \Pr[x < y | n \geq 1  , m \geq 1] \nonumber \\
& = \int_{y=0}^{d_{\max}} \int_{x=0}^{y} \! \sum\limits_{n=1}^{\infty} {\sum\limits_{m=1}^{\infty} {f_{X_{(1)},n_I} \left( X_{(1)} = x , n_I = n | n \geq 1 \right) f_{Y_{(1)},n_o} \left( Y_{(1)} = y , n_o = m | m \geq 1 \right)}} \, \mathrm{d}x \mathrm{d}y \nonumber \\
&\stackrel{\text{\eqref{eq: jointPDF-Lemma1}}}{=}
\frac{4 \lambda_I \lambda_o A_{d_{\max}}^{2}}{d_{\max}^{4}}
\int_{y=0}^{d_{\max}} \int_{x=0}^{y} \!
\frac{xye^{- \left( \lambda_I + \lambda_o \right) A_{d_{\max}}}}{\left( 1 - e^{-\lambda_I A_{d_{\max}}} \right)\left( 1 - e^{-\lambda_o A_{d_{\max}}} \right)} \nonumber \\
& \hspace{43mm} \times \sum\limits_{n=1}^{\infty} {\frac{\biggl( \left( 1 - \frac{x^2}{d_{\max}^{2}} \right) \lambda_I A_{d_{\max}} \biggr)^{n-1}}{\left( n - 1\right)!}} \sum\limits_{m=1}^{\infty} {\frac{\biggl( \left( 1 - \frac{y^2}{d_{\max}^{2}} \right) \lambda_o A_{d_{\max}} \biggr)^{m-1}}{\left( m - 1\right)!}} \, \mathrm{d}x \mathrm{d}y \nonumber \\
&\stackrel{(\star)}{=}
\frac{4 \lambda_I \lambda_o A_{d_{\max}}^{2}}{d_{\max}^{4}\left( 1 - e^{-\lambda_I A_{d_{\max}}} \right)\left( 1 - e^{-\lambda_o A_{d_{\max}}} \right)}
\int_{y=0}^{d_{\max}} \int_{x=0}^{y} \!
e^{- \left( \lambda_I + \lambda_o \right) A_{d_{\max}}} e^{\left( 1 - x^2/d_{\max}^{2} \right) \lambda_I A_{d_{\max}}}
e^{\left( 1 - y^2/d_{\max}^{2} \right) \lambda_o A_{d_{\max}}} xy\, \mathrm{d}x \mathrm{d}y \nonumber \\
& = \frac{4 \lambda_I \lambda_o A_{d_{\max}}^{2}}{d_{\max}^{4}\left( 1 - e^{-\lambda_I A_{d_{\max}}} \right)\left( 1 - e^{-\lambda_o A_{d_{\max}}} \right)}
\int_{y=0}^{d_{\max}} \!
y e^{- \lambda_o A_{d_{\max}} y^2 \big/ d_{\max}^{2}}
\int_{x=0}^{y} x e^{- \lambda_I A_{d_{\max}} x^2 \big/ d_{\max}^{2}} \, \mathrm{d}x \mathrm{d}y \nonumber \\
& = \frac{\lambda_o}{\left( 1 - e^{-\lambda_I A_{d_{\max}}} \right)\left( 1 - e^{-\lambda_o A_{d_{\max}}} \right)}
\left( \frac{1 - e^{-\lambda_o A_{d_{\max}}}}{\lambda_o} -
\frac{1 - e^{-\left( \lambda_o + \lambda_I \right) A_{d_{\max}}}}{\lambda_o + \lambda_I}  \right) \:.
\end{align}
\hrulefill
\end{figure*}

\section*{Appendix~B: \break Throughput Analysis of TDMA}\label{appen: TDMA-throughput-analaysis}
Consider a network of area $A$, TDMA-based channel access, and $1+n_t$ links including the typical link, where $n_t$ is a Poisson random variable with mean $A \lambda_t$. Also, assume that the intended receiver of each transmitter $i$ is located at distance $0 < L_i \leq d_{\max}$ at the cone where the transmitter's signal is pointed. Having a natural assumption of the independence of the lengths of different links, $\left\{L_i\right\}_{i=1}^{1+n_t}$ become i.i.d random variables with density function $f_{L} \left( \ell \right) = 2 \ell / d_{\max}^{2}$. Let $z_{L_i}$ be a binary random variable taking 1 if and only if link $i$ has the LoS condition (no blockage). As there is no concurrent transmissions in TDMA, the success probability for TDMA given $L_i$ and $n_t$ is equal to having no obstacle on link $i$, which occurs with probability $\Pr[z_{L_i}=1~|L_i,n_t] = e^{- \lambda_o A_{L_i}}$, see Fig.~\ref{fig: IntRegion}.
In long term, TDMA scheduler allocates only $1/(1+n_t)$ shares of the total resources to every link. Assuming transmission of one packet per slot, the MAC throughput of each link $i$ in TDMA, denoted by $r_{_{\text{TDMA}}}$, is
\begin{align}\label{eq: MACThroughpuTDMA1}
r_{_{\text{TDMA}}} &=  \sum\limits_{n_t=0}^{\infty} \frac{e^{-A \lambda_t}}{\left( 1 + n_t \right) } \frac{\left( A \lambda_t \right)^ {n_t}}{n_t !} \int_{\ell_{i}=0}^{d_{\max}} \! e^{- \lambda_o \theta_c \ell_{i}^{2}/2} \frac{2\ell_i}{d_{\max}^{2}}\, \mathrm{d}\ell_{i} \nonumber \\
&=  \left( \frac{1 - e^{-A \lambda_t}}{A \lambda_t} \right) \frac{2}{d_{\max}^{2}} \left( \frac{1-e^{-\lambda_o A_{d_{\max}}}}{\lambda_o \theta_c} \right) \:.
\end{align}

Recalling $A_{d_{\max}} = \theta_c d_{\max}^{2}/2$, \eqref{eq: MACThroughpuTDMA1} simplifies to~\eqref{eq: TDMALinkThroughput}. To find the area spectral efficiency of TDMA scheduler, we assume that $z_{L_i}$ and $z_{L_j}$ are independent\footnote{This independence means that the event of having obstacle on the path between different transmitter-receiver pairs are independent. Still, we have correlated LoS conditions (angular correlation) on the channels between different transmitters and a common receiver.} for all $L_i$, $L_j$, $i$, and $j$, where $j \neq i$. The area spectral efficiency of TDMA, denoted by ${\mathrm{ASE}}_{\text{TDMA}}$, is derived in~\eqref{eq: ASEThroughpuTDMA}, where ${\mathrm{ASE}}_{\text{TDMA} \mid n_t}$ is the area spectral efficiency of TDMA given $n_t$ and $f_{L_1, \ldots, L_{1+n_t}}\left( \ell_1, \ldots, \ell_{1+n_t} \right)$ is joint distribution of the links lengths. This concludes the proof.
\begin{figure*}[!t]
\normalsize
\begin{align}\label{eq: ASEThroughpuTDMA}
{\mathrm{ASE}}_{\text{TDMA}} &= \frac{1}{A} \sum\limits_{n_t=0}^{\infty} e^{-A \lambda_t} \frac{\left( A \lambda_t \right)^ {n_t}}{n_t !} {\mathrm{ASE}}_{\text{TDMA} \mid n_t}  \nonumber \\
&= \frac{1}{A} \sum\limits_{n_t=0}^{\infty} \frac{e^{-A \lambda_t} }{1+n_t}\frac{\left( A \lambda_t \right)^ {n_t}}{n_t !} \hspace{-1mm} \int_{\ell_{1}=0}^{d_{\max}} \hspace{-1mm} \cdots \int_{\ell_{1+n_t}=0}^{d_{\max}}\!  \sum\limits_{i = 1}^{1+n_t} \Pr[z_{\ell_i}=1~|\ell_1, \ldots, \ell_{1+n_t}, n_t] \, f_{L_1, \ldots, L_{1+n_t}}\hspace{-0.5mm} \left( \ell_1, \ldots, \ell_{1+n_t} \right)
 \mathrm{d}\ell_{1} \ldots \mathrm{d}\ell_{1+n_t}
 \nonumber \\
&= \frac{1}{A} \sum\limits_{n_t=0}^{\infty} \frac{e^{-A \lambda_t} }{1+n_t}\frac{\left( A \lambda_t \right)^ {n_t}}{n_t !} \sum\limits_{i = 1}^{1+n_t} \left( \left( \int_{\ell_{i}=0}^{d_{\max}} \! \Pr[z_{\ell_i}=1~|\ell_{i}, n_t] \, f_{L_{i}} \hspace{-0.5mm} \left( \ell_i \right) \, \mathrm{d}\ell_{i}  \right) \prod_{j = 1 \atop j \neq i}^{1+n_t} \int_{\ell_{j}=0}^{d_{\max}} \! f_{L_j} \left( \ell_j \right) \, \mathrm{d}\ell_{j} \right )  \nonumber \\
&= \frac{1}{A} \sum\limits_{n_t=0}^{\infty}  \frac{e^{-A \lambda_t} }{1+n_t}\frac{\left( A \lambda_t \right)^ {n_t}}{n_t !} \sum\limits_{i = 1}^{1+n_t} \int_{\ell_{i}=0}^{d_{\max}} \! e^{- \lambda_o \theta_c \ell_{i}^{2}/2} \frac{2\ell_i}{d_{\max}^{2}} \, \mathrm{d}\ell_{i}  \nonumber \\
&= \frac{1}{A} \sum\limits_{n_t=0}^{\infty}  \frac{e^{-A \lambda_t} }{1+n_t}\frac{\left( A \lambda_t \right)^ {n_t}}{n_t !} \left( 1 + n_t \right)  \left( \frac{1-e^{-\lambda_o A_{d_{\max}}}}{\lambda_o A_{d_{\max}}} \right) =  \frac{1-e^{-\lambda_o A_{d_{\max}}}}{A\lambda_o A_{d_{\max} }} \:.
\end{align}
\hrulefill
\end{figure*}

\bibliographystyle{IEEEtran}
\bibliography{References}

\begin{thebibliography}{10}
\providecommand{\url}[1]{#1}
\csname url@samestyle\endcsname
\providecommand{\newblock}{\relax}
\providecommand{\bibinfo}[2]{#2}
\providecommand{\BIBentrySTDinterwordspacing}{\spaceskip=0pt\relax}
\providecommand{\BIBentryALTinterwordstretchfactor}{4}
\providecommand{\BIBentryALTinterwordspacing}{\spaceskip=\fontdimen2\font plus
\BIBentryALTinterwordstretchfactor\fontdimen3\font minus
  \fontdimen4\font\relax}
\providecommand{\BIBforeignlanguage}[2]{{%
\expandafter\ifx\csname l@#1\endcsname\relax
\typeout{** WARNING: IEEEtran.bst: No hyphenation pattern has been}%
\typeout{** loaded for the language `#1'. Using the pattern for}%
\typeout{** the default language instead.}%
\else
\language=\csname l@#1\endcsname
\fi
#2}}
\providecommand{\BIBdecl}{\relax}
\BIBdecl

\bibitem{ECMA387}
{ECMA-TC48, ECMA standard 387}, ``High rate 60 {GHz} {PHY}, {MAC} and {HDMI}
  {PAL},'' Dec. 2008.

\bibitem{802_15_3c}
``{IEEE} 802.15.3c {P}art 15.3: {W}ireless medium access control ({MAC}) and
  physical layer ({PHY}) specifications for high rate wireless personal area
  networks ({WPANs}) amendment 2: {M}illimeter-wave-based alternative physical
  layer extension,'' Oct. 2009.

\bibitem{802_11ad}
``{IEEE} 802.11ad. {P}art 11: {W}ireless {LAN} medium access control ({MAC})
  and physical layer ({PHY}) specifications - amendment 3: {E}nhancements for
  very high throughput in the 60 {GHz} band,'' Dec. 2012.

\bibitem{FCC2}
\BIBentryALTinterwordspacing
{Federal Communications Commission}, ``{FCC-177},'' Jan. 2015. [Online].
  Available: \url{http://apps.fcc.gov/ecfs/proceeding/view?name=14-177}
\BIBentrySTDinterwordspacing

\bibitem{Ofcom1}
\BIBentryALTinterwordspacing
{Ofcom}, ``Spectrum above {6GHz} for future mobile communications,'' Feb. 2015.
  [Online]. Available:
  \url{http://stakeholders.ofcom.org.uk/binaries/consultations/above-6ghz/summary/spectrum_above_6_GHz_CFI.pdf}
\BIBentrySTDinterwordspacing

\bibitem{Rappaport2013Millimeter}
T.~Rappaport, S.~Sun, R.~Mayzus, H.~Zhao, Y.~Azar, K.~Wang, G.~Wong, J.~Schulz,
  M.~Samimi, and F.~Gutierrez, ``Millimeter wave mobile communications for {5G}
  cellular: {I}t will work!'' \emph{{IEEE} Access}, vol.~1, pp. 335--349, May
  2013.

\bibitem{Andrews2014What}
J.~G. Andrews, S.~Buzzi, W.~Choi, S.~Hanly, A.~Lozano, A.~C. Soong, and J.~C.
  Zhang, ``What will {5G} be?'' \emph{{IEEE} J. Select. Areas Commun.},
  vol.~32, no.~6, pp. 1065--1082, Jun. 2014.

\bibitem{osseiran2014scenarios}
A.~Osseiran, F.~Boccardi, V.~Braun, K.~Kusume, P.~Marsch, M.~Maternia,
  O.~Queseth, M.~Schellmann, H.~Schotten, H.~Taoka \emph{et~al.}, ``Scenarios
  for {5G} mobile and wireless communications: the vision of the {METIS}
  project,'' \emph{{IEEE} Commun. Mag.}, vol.~52, no.~5, pp. 26--35, May 2014.

\bibitem{boccardi2014Five}
F.~Boccardi, R.~Heath, A.~Lozano, T.~L. Marzetta, and P.~Popovski, ``Five
  disruptive technology directions for {5G},'' \emph{{IEEE} Commun. Mag.},
  vol.~52, no.~2, pp. 74--80, Feb. 2014.

\bibitem{shokri2015mmWavecellular}
H.~Shokri-Ghadikolaei, C.~Fischione, G.~Fodor, P.~Popovski, and M.~Zorzi,
  ``Millimeter wave cellular networks: {A} {MAC} layer perspective,''
  \emph{{IEEE} Trans. Commun.}, vol.~63, no.~10, pp. 3437--3458, Oct. 2015.

\bibitem{Niu2015Survey}
Y.~Niu, Y.~Li, D.~Jin, L.~Su, and A.~Vasilakos, ``A survey of millimeter wave
  communications {(mmWave)} for {5G}: {O}pportunities and challenges,''
  \emph{Wireless Networks}, pp. 1--20, Apr. 2015.

\bibitem{Rangan2014Millimeter}
S.~Rangan, T.~Rappaport, and E.~Erkip, ``Millimeter wave cellular wireless
  networks: {P}otentials and challenges,'' \emph{Proc. {IEEE}}, vol. 102,
  no.~3, pp. 366--385, Mar. 2014.

\bibitem{Shokri2015mmWaveWPAN}
H.~Shokri-Ghadikolaei, C.~Fischione, P.~Popovski, and M.~Zorzi, ``Design
  aspects of short range millimeter wave wireless networks: {A} {MAC} layer
  perspective,'' \emph{{IEEE} Netw.}, 2015, to appear.

\bibitem{nelson1985spatial}
R.~Nelson and L.~Kleinrock, ``Spatial {TDMA}: {A} collision-free multihop
  channel access protocol,'' \emph{{IEEE} Trans. Commun.}, vol.~33, no.~9, pp.
  934--944, Sept. 1985.

\bibitem{an2008directional}
X.~An and R.~Hekmat, ``Directional {MAC} protocol for millimeter wave based
  wireless personal area networks,'' in \emph{Proc. {IEEE} Vehicular Technology
  Conference (VTC Spring)}, 2008, pp. 1636--1640.

\bibitem{Shokri2015Beam}
H.~Shokri-Ghadikolaei, L.~Gkatzikis, and C.~Fischione, ``Beam-searching and
  transmission scheduling in millimeter wave communications,'' in \emph{Proc.
  {IEEE} International Conference on Communications (ICC)}, 2015, pp.
  1292--1297.

\bibitem{bjorklund2003resource}
P.~Bjorklund, P.~Varbrand, and D.~Yuan, ``Resource optimization of spatial tdma
  in ad hoc radio networks: {A} column generation approach,'' in \emph{Proc.
  {IEEE} International Conference on Computer Communications (INFOCOM)}, 2003,
  pp. 818--824.

\bibitem{gronkvist2001comparison}
J.~Gr{\"o}nkvist and A.~Hansson, ``Comparison between graph-based and
  interference-based stdma scheduling,'' in \emph{Proc. {ACM} international
  symposium on Mobile ad hoc networking \& computing (MobiHoc)}, 2001, pp.
  255--258.

\bibitem{Nitsche2014IEEE}
T.~Nitsche, C.~Cordeiro, A.~B. Flores, E.~W. Knightly, E.~Perahia, and J.~C.
  Widmer, ``{IEEE 802.11ad}: {D}irectional {60~GHz} communication for
  multi-{G}igabit-per-second {Wi-Fi},'' \emph{{IEEE} Commun. Mag.}, vol.~52,
  no.~12, pp. 132--141, Dec. 2014.

\bibitem{ramanathan1997unified}
S.~Ramanathan, ``A unified framework and algorithm for {(T/F/C) DMA} channel
  assignment in wireless networks,'' in \emph{Proc. {IEEE} International
  Conference on Computer Communications (INFOCOM)}, 1997, pp. 900--907.

\bibitem{rhee2008z}
I.~Rhee, A.~Warrier, M.~Aia, J.~Min, and M.~L. Sichitiu, ``{Z-MAC}: {A} hybrid
  {MAC} for wireless sensor networks,'' \emph{{IEEE/ACM} Trans. Netw.},
  vol.~16, no.~3, pp. 511--524, Jun. 2008.

\bibitem{son2012frame}
I.~K. Son, S.~Mao, M.~X. Gong, and Y.~Li, ``On frame-based scheduling for
  directional {mmWave} {WPANs},'' in \emph{Proc. {IEEE} International
  Conference on Computer Communications (INFOCOM)}, 2012, pp. 2149--2157.

\bibitem{niu2015exploiting}
Y.~Niu, C.~Gao, Y.~Li, L.~Su, D.~Jin, and A.~Vasilakos, ``Exploiting
  device-to-device communications in joint scheduling of access and backhaul
  for {mmWave} small cells,'' \emph{{IEEE} J. Sel. Areas Commun.}, vol.~33,
  no.~10, pp. 2052--2069, Oct. 2015.

\bibitem{magistretti2014802}
E.~Magistretti, O.~Gurewitz, and E.~W. Knightly, ``802.11ec: {C}ollision
  avoidance without control messages,'' \emph{{IEEE/ACM} Trans. Netw.},
  vol.~22, no.~6, pp. 1845--1858, Dec. 2014.

\bibitem{ephremides1982analysis}
A.~Ephremides and O.~A. Mowafi, ``Analysis of a hybrid access scheme for
  buffered users-probabilistic time division,'' \emph{{IEEE} Trans. Software
  Eng.}, no.~1, pp. 52--61, Jan. 1982.

\bibitem{rios1985hybrid}
M.~Rios and N.~D. Georganas, ``A hybrid multiple-access protocol for data and
  voice-packet over local area networks,'' \emph{{IEEE} Trans. Comput.}, vol.
  100, no.~1, pp. 90--94, Jan. 1985.

\bibitem{van2003adaptive}
T.~Van~Dam and K.~Langendoen, ``An adaptive energy-efficient {MAC} protocol for
  wireless sensor networks,'' in \emph{Proc. {ACM} international conference on
  mbedded networked sensor systems (SenSys)}, 2003, pp. 171--180.

\bibitem{ye2004medium}
W.~Ye, J.~Heidemann, and D.~Estrin, ``Medium access control with coordinated
  adaptive sleeping for wireless sensor networks,'' \emph{{IEEE/ACM} Trans.
  Netw.}, vol.~12, no.~3, pp. 493--506, Jun. 2004.

\bibitem{vuran2007mac}
M.~C. Vuran and I.~F. Akyildiz, ``{A-MAC}: {A}daptive medium access control for
  next generation wireless terminals,'' \emph{{IEEE/ACM} Trans. Netw.},
  vol.~15, no.~3, pp. 574--587, Jun. 2007.

\bibitem{Singh2011Interference}
S.~Singh, R.~Mudumbai, and U.~Madhow, ``Interference analysis for highly
  directional 60-{GHz} mesh networks: {T}he case for rethinking medium access
  control,'' \emph{{IEEE/ACM} Trans. Netw.}, vol.~19, no.~5, pp. 1513--1527,
  Oct. 2011.

\bibitem{Qiao2015D2D}
J.~Qiao, X.~Shen, J.~Mark, Q.~Shen, Y.~He, and L.~Lei, ``Enabling
  device-to-device communications in millimeter-wave {(5G)} cellular
  networks,'' \emph{{IEEE} Commun. Mag.}, vol.~53, no.~1, pp. 209--215, Jan.
  2015.

\bibitem{Niu2015Blockage}
Y.~Niu, Y.~Li, D.~Jin, L.~Su, and D.~Wu, ``Blockage robust and efficient
  scheduling for directional {mmWave WPANs},'' \emph{{IEEE} Trans. Veh.
  Technol.}, vol.~64, no.~2, pp. 728--742, Feb. 2015.

\bibitem{Son2012on}
I.~Son, S.~Mao, M.~Gong, and Y.~Li, ``On frame-based scheduling for directional
  {mmWave} {WPANs},'' in \emph{Proc. {IEEE} International Conference on
  Computer Communications (INFOCOM)}, 2012, pp. 2149--2157.

\bibitem{Park2009Analysis}
M.~Park and P.~Gopalakrishnan, ``Analysis on spatial reuse and interference in
  60-{GHz} wireless networks,'' \emph{{IEEE} J. Sel. Areas Commun.}, vol.~27,
  no.~8, pp. 1443--1452, Oct. 2009.

\bibitem{Qiao2012STDMA}
J.~Qiao, L.~X. Cai, X.~Shen, and J.~Mark, ``{STDMA}-based scheduling algorithm
  for concurrent transmissions in directional millimeter wave networks,'' in
  \emph{Proc. {IEEE} International Conference on Communications (ICC)}, 2012,
  pp. 5221--5225.

\bibitem{Sum2009virtual}
C.~Sum, Z.~Lan, R.~Funada, J.~Wang, T.~Baykas, M.~A. Rahman, and H.~Harada,
  ``Virtual time-slot allocation scheme for throughput enhancement in a
  millimeter-wave multi-{Gbps} {WPAN} system,'' \emph{{IEEE} J. Sel. Areas
  Commun.}, vol.~27, no.~8, pp. 1379--1389, Oct. 2009.

\bibitem{di2014stochastic}
M.~Di~Renzo, ``Stochastic geometry modeling and analysis of multi-tier
  millimeter wave cellular networks,'' \emph{{IEEE} Trans. Wireless Commun.},
  vol.~14, no.~9, pp. 5038--5057, Sept. 2015.

\bibitem{park2015TractableResource}
J.~Park, S.-L. Kim, and J.~Zander, ``Tractable resource management in
  millimeter-wave overlaid ultra-dense cellular networks,'' \emph{arXiv
  preprint arXiv:1507.04658}, 2015.

\bibitem{lu2015stochastic}
W.~Lu and M.~{Di Renzo}, ``Stochastic geometry modeling of {mmWave} cellular
  networks: {A}nalysis and experimental validation,'' in \emph{Proc. {IEEE}
  International Workshop on Measurements Networking}, Oct. 2015, pp. 1--4.

\bibitem{Singh2015TractableModel}
S.~Singh, M.~N. Kulkarni, A.~Ghosh, and J.~G. Andrews, ``Tractable model for
  rate in self-backhauled millimeter wave cellular networks,'' \emph{{IEEE} J.
  Sel. Areas Commun.}, vol.~33, no.~10, pp. 2196--2211, Oct. 2015.

\bibitem{TBai2014Coverage}
T.~Bai and R.~Heath, ``Coverage and rate analysis for millimeter wave cellular
  networks,'' \emph{{IEEE} Trans. Wireless Commun.}, vol.~14, no.~2, pp.
  1100--1114, Feb. 2015.

\bibitem{hunter2008transmission}
A.~M. Hunter, J.~G. Andrews, and S.~Weber, ``Transmission capacity of ad hoc
  networks with spatial diversity,'' \emph{{IEEE} Trans. Wireless Commun.},
  vol.~7, no.~12, pp. 5058--5071, Dec. 2008.

\bibitem{wildman2014joint}
J.~Wildman, P.~H. Nardelli, M.~Latva-aho, and S.~Weber, ``On the joint impact
  of beamwidth and orientation error on throughput in wireless directional
  poisson networks,'' \emph{{IEEE} Trans. Wireless Commun.}, vol.~13, no.~12,
  pp. 7072--7085, Dec. 2014.

\bibitem{shokri2015MillimeterGlobecom}
H.~Shokri-Ghadikolaei and C.~Fischione, ``Millimeter wave ad hoc networks:
  {N}oise-limited or interference-limited?'' in \emph{Proc. {IEEE} Global
  Telecommunications Conference (GLOBECOM) Workshops}, 2015.

\bibitem{pyo2009throughput}
C.~W. Pyo and H.~Harada, ``Throughput analysis and improvement of hybrid
  multiple access in {IEEE 802.15.3c} mm-wave {WPAN},'' \emph{{IEEE} J. Sel.
  Areas Commun.}, vol.~27, no.~8, pp. 1414--1424, Oct. 2009.

\bibitem{Haenggi2013Stochastic}
M.~Haenggi, \emph{Stochastic Geometry for Wireless Networks}.\hskip 1em plus
  0.5em minus 0.4em\relax Cambridge University Press, 2013.

\bibitem{kleinrock1975packet}
L.~Kleinrock and F.~A. Tobagi, ``Packet switching in radio channels: {Part
  I}--carrier sense multiple-access modes and their throughput-delay
  characteristics,'' \emph{{IEEE} Trans. Commun.}, vol.~23, no.~12, pp.
  1400--1416, Dec. 1975.

\bibitem{singh2009blockage}
S.~Singh, F.~Ziliotto, U.~Madhow, E.~Belding, and M.~Rodwell, ``Blockage and
  directivity in 60 {GHz} wireless personal area networks: {F}rom cross-layer
  model to multihop {MAC} design,'' \emph{{IEEE} J. Sel. Areas Commun.},
  vol.~27, no.~8, pp. 1400--1413, Oct. 2009.

\bibitem{Rappaport2015wideband}
T.~S. Rappaport, G.~R. MacCartney, M.~K. Samimi, and S.~Sun, ``Wideband
  millimeter-wave propagation measurements and channel models for future
  wireless communication system design,'' \emph{{IEEE} Trans. Commun.},
  vol.~63, no.~9, pp. 3029--3056, Sept. 2015.

\bibitem{gupta2000capacity}
P.~Gupta and P.~R. Kumar, ``The capacity of wireless networks,'' \emph{{IEEE}
  Trans. Inform. Theory}, vol.~46, no.~2, pp. 388--404, Mar. 2000.

\bibitem{xu2002effective}
K.~Xu, M.~Gerla, and S.~Bae, ``How effective is the {IEEE} 802.11 {RTS/CTS}
  handshake in ad hoc networks?'' in \emph{Proc. {IEEE} Global
  Telecommunications Conference (GLOBECOM)}, 2002, pp. 72--76.

\bibitem{iyer2009right}
A.~Iyer, C.~Rosenberg, and A.~Karnik, ``What is the right model for wireless
  channel interference?'' \emph{{IEEE} Trans. Wireless Commun.}, vol.~8, no.~5,
  pp. 2662--2671, May 2009.

\bibitem{cardieri2010modeling}
P.~Cardieri, ``Modeling interference in wireless ad hoc networks,''
  \emph{{IEEE} Commun. Surveys Tuts.}, vol.~12, no.~4, pp. 551--572, Fourth
  Quarter 2010.

\bibitem{Shokri2015OntheAccuracy}
H.~Shokri-Ghadikolaei, C.~Fischione, and E.~Modiano, ``On the accuracy of
  interference models in wireless communications,'' KTH Royal Institute of
  Technology, Tech. Rep., Sept. 2015.

\bibitem{Shokri2015WhatIs}
H.~Shokri-Ghadikolaei, X.~Jiang, C.~Fischione, and P.~Zhibo, ``What is the
  right interference model in millimeter wave networks?'' KTH Royal Institute
  of Technology, Tech. Rep., Nov. 2015.

\bibitem{TBai2014Blockage}
T.~Bai, R.~Vaze, and R.~Heath, ``Analysis of blockage effects on urban cellular
  networks,'' \emph{{IEEE} Trans. Wireless Commun.}, vol.~13, no.~9, pp.
  5070--5083, Sept. 2014.

\bibitem{Akdeniz2014MillimeterWave}
M.~Akdeniz, Y.~Liu, M.~Samimi, S.~Sun, S.~Rangan, T.~Rappaport, and E.~Erkip,
  ``Millimeter wave channel modeling and cellular capacity evaluation,''
  \emph{{IEEE} J. Sel. Areas Commun.}, vol.~32, no.~6, pp. 1164--1179, Jun.
  2014.

\bibitem{goldsmith2003capacity}
A.~Goldsmith, S.~A. Jafar, N.~Jindal, and S.~Vishwanath, ``Capacity limits of
  {MIMO} channels,'' \emph{{IEEE} J. Sel. Areas Commun.}, vol.~21, no.~5, pp.
  684--702, Jun. 2003.

\bibitem{Zhao201328Ghz}
H.~Zhao, R.~Mayzus, S.~Sun, M.~Samimi, J.~K. Schulz, Y.~Azar, K.~Wang, G.~N.
  Wong, F.~Gutierrez, and T.~S. Rappaport, ``28 {GHz} millimeter wave cellular
  communication measurements for reflection and penetration loss in and around
  buildings in {New York} city,'' in \emph{Proc. {IEEE} International
  Conference on Communications (ICC)}, 2013, pp. 5163--5167.

\bibitem{park2009generalized}
P.~Park, P.~Di~Marco, P.~Soldati, C.~Fischione, and K.~H. Johansson, ``A
  generalized {Markov} chain model for effective analysis of slotted {IEEE
  802.15.4},'' in \emph{Proc. {IEEE} Mobile Adhoc and Sensor Systems (MASS)},
  2009, pp. 130--139.

\bibitem{pollin2008performance}
S.~Pollin, M.~Ergen, S.~Ergen, B.~Bougard, L.~Der~Perre, I.~Moerman, A.~Bahai,
  P.~Varaiya, and F.~Catthoor, ``Performance analysis of slotted carrier sense
  {IEEE 802.15.4} medium access layer,'' \emph{{IEEE} Trans. Wireless Commun.},
  vol.~7, no.~9, pp. 3359--3371, Sept. 2008.

\bibitem{bianchi2000performance}
G.~Bianchi, ``Performance analysis of the {IEEE 802.11} distributed
  coordination function,'' \emph{{IEEE} J. Sel. Areas Commun.}, vol.~18, no.~3,
  pp. 535--547, Mar. 2000.

\bibitem{hui2005unified}
J.~Hui and M.~Devetsikiotis, ``A unified model for the performance analysis of
  {IEEE} 802.11e {EDCA},'' \emph{{IEEE} Trans. Commun.}, vol.~53, no.~9, pp.
  1498--1510, Sept. 2005.

\bibitem{malone2007modeling}
D.~Malone, K.~Duffy, and D.~Leith, ``Modeling the 802.11 distributed
  coordination function in nonsaturated heterogeneous conditions,''
  \emph{{IEEE/ACM} Trans. Netw.}, vol.~15, no.~1, pp. 159--172, Feb. 2007.

\bibitem{garetto2008modeling}
M.~Garetto, T.~Salonidis, and E.~W. Knightly, ``Modeling per-flow throughput
  and capturing starvation in {CSMA} multi-hop wireless networks,''
  \emph{{IEEE/ACM} Trans. Netw.}, vol.~16, no.~4, pp. 864--877, Aug. 2008.

\bibitem{jang2012ieee}
B.~Jang and M.~L. Sichitiu, ``{IEEE} 802.11 saturation throughput analysis in
  the presence of hidden terminals,'' \emph{{IEEE/ACM} Trans. Netw.}, vol.~20,
  no.~2, pp. 557--570, Apr. 2012.

\bibitem{Yang2003Delay}
Y.~Yang and T.~Yum, ``Delay distributions of slotted {ALOHA} and {CSMA},''
  \emph{{IEEE} Trans. Commun.}, vol.~51, no.~11, pp. 1846--1857, Nov. 2003.

\bibitem{benvenuto2011principles}
N.~Benvenuto and M.~Zorzi, \emph{Principles of communications Networks and
  Systems}.\hskip 1em plus 0.5em minus 0.4em\relax Wiley Online Library, 2011.

\bibitem{Shokri2015Licentiate}
H.~Shokri-Ghadikolaei, ``Fundamentals of medium access control design for
  millimeter wave networks,'' Licentiate Thesis, KTH, Royal Institute of
  Technology, Sept. 2015.

\bibitem{david1970order}
H.~A. David and H.~N. Nagaraja, \emph{Order statistics}.\hskip 1em plus 0.5em
  minus 0.4em\relax Wiley Online Library, 2003, third Edition.

\bibitem{johnson2005univariate}
N.~L. Johnson, A.~W. Kemp, and S.~Kotz, \emph{Univariate discrete
  distributions}.\hskip 1em plus 0.5em minus 0.4em\relax John Wiley \& Sons,
  2005, vol. 444.

\end{thebibliography}

\end{document}